\documentclass[review]{elsarticle}
\journal{Journal of \LaTeX\ Templates}

\usepackage[a4paper, total={6in, 8in}]{geometry}
\usepackage{amsmath}
\usepackage{pgfplots}
\usepackage{tikz}

\usepgfplotslibrary{fillbetween,groupplots,external}
\usetikzlibrary{backgrounds,calc,arrows,positioning,matrix,external,pgfplots.external}

\tikzset{framed, tight background}
\pgfplotsset{
scaled y ticks=false,
label style={font=\footnotesize},
tick label style={font=\footnotesize},
}

\usepackage{graphicx}
\usepackage{epstopdf, epsfig}
\usepackage{amssymb}
\usepackage{pgffor}
\usepackage{xcolor}
\usepackage{stix}
\usepackage{mathtools}

\usepackage{nomencl}
\usepackage{multicol}
\usepackage[nopostdot,nogroupskip,nonumberlist,toc, style=tree, section, shortcuts]{glossaries}
\usepackage{subcaption}
\usepackage{glossaries}

\newacronym{BC}{BC}{Boundary Condition}
\newacronym{CAA}{CAA}{Computational Aero-Acoustics}
\newacronym{CFD}{CFD}{Computational Fluid Dynamics}
\newacronym{CFL}{CFL}{Courant-Friedrichs-Lewy}
\newacronym{DNS}{DNS}{Direct Numerical Simulation}
\newacronym{EBU}{EBU}{Eddy Break Up}
\newacronym{EVP}{EVP}{Eigenvalue Problem}
\newacronym{FDF}{FDF}{Flame Describing Function}
\newacronym{FE}{FE}{Finite Element}
\newacronym{FELiCS}{FELiCS}{Finite Element Linearized Combustion Solver}
\newacronym{FTF}{FTF}{Flame Transfer Function}
\newacronym{GEVP}{GEVP}{Generalized Eigenvalue Problem}
\newacronym{KH}{KH}{Kelvin-Helmholtz}
\newacronym{LES}{LES}{Large Eddy Simulation}
\newacronym{LHS}{LHS}{left hand side}
\newacronym{LNSE}{LNSE}{lineaerized Navier-Stokes equations}
\newacronym{LSA}{LSA}{Linear Stability Analysis}
\newacronym{PVC}{PVC}{Precessing Vortex Core}
\newacronym{RA}{RA}{Resolvent Analysis}
\newacronym{RANS}{RANS}{Reynolds-averaged Navier Stokes}
\newacronym{RHS}{RHS}{right hand side}
\newacronym{RST}{RST}{Reynolds-Stress}
\newacronym{SGS}{SGS}{Subgrid Scale}
\newacronym{SPL}{SPL}{Sound Pressure Level}
\newacronym{SPOD}{SPOD}{Spectral Proper Orthogonal Decomposition}
\newacronym{TFC}{TFC}{Turbulent Flame-speed Closure}
\newacronym{TKE}{TKE}{Turbulent-Kinetic-Energy}

\newacronym{felics}{FELiCS}{Finite Element Linear Combustion Solver}
\newacronym{fenics}{FEniCS}{Finite Element Computational Software}
\usepackage{ulem}

\usepackage{numcompress}
\usepackage{etoolbox}

\usepackage{lipsum}
\newsavebox{\arrangebox}

\usepackage{scalerel}
\usepackage{stackengine,wasysym}

\newcommand\reallywidetilde[1]{\ThisStyle{%
  \setbox0=\hbox{$\SavedStyle#1$}%
  \stackengine{-.1\LMpt}{$\SavedStyle#1$}{%
    \stretchto{\scaleto{\SavedStyle\mkern.2mu\AC}{.5150\wd0}}{.6\ht0}%
  }{O}{c}{F}{T}{S}%
}}

\usepackage{bm}
\usepackage{physics}
\newcommand{\tlk}[1]{\textcolor{black}{#1}}
\newcommand{\filePrefix}{fig}
\newcommand{\tikzpath}{epsFigs/}
\newcommand{\figureMode}{2}

\newcommand{\externalizedFigurePath}{AtThisPointThisIsARandomString}

\def\firstoftwo#1#2{#1}
\def\secondoftwo#1#2{#2}
\def\iffileexist#1{%
  \expandafter\ifx\expandafter&\pdffilesize{#1}&%
    \expandafter\secondoftwo
  \else
    \expandafter\firstoftwo
  \fi
}
\def\iffileempty#1{%
  \ifnum0\pdffilesize{#1}>0
    \expandafter\secondoftwo
  \else
    \expandafter\firstoftwo
  \fi
}
\if\figureMode3
    \usetikzlibrary{external}
    \newcounter{externalizedFigure}
    \newcounter{externalizedSubFigure}
    \tikzexternaldisable
\fi
\if\figureMode4
    \newcounter{externalizedFigure}
    \newcounter{externalizedSubFigure}
\fi

\newcommand{\includeFigureOwn}[1]{
    \if\figureMode1
        \includegraphics[width=\textwidth]{example-image-a}
    \fi
    \if\figureMode2
        \input{#1}
    \fi
    \if\figureMode3
        \tikzexternalenable
        \stepcounter{externalizedFigure}
        \renewcommand{\externalizedFigurePath}{\filePrefix\theexternalizedFigure}
        \tikzsetnextfilename{\externalizedFigurePath}
        \input{#1}
        \tikzexternaldisable
    \fi
    \if\figureMode4
       \stepcounter{externalizedFigure}
       \includegraphics[scale=1]{\tikzpath\filePrefix\theexternalizedFigure.pdf}
    \fi
    \if\figureMode5
       \stepcounter{externalizedFigure}
       \scalebox{0}{
            \input{#1}
        }
       \includegraphics[scale=1]{\tikzpath\filePrefix\theexternalizedFigure}
    \fi
}

\newcommand{\includeFirstSubFigureOwn}[1]{\if\figureMode3
        \setcounter{externalizedSubFigure}{0}
        \stepcounter{externalizedFigure}
    \fi
    \if\figureMode4
       \setcounter{externalizedSubFigure}{0}
       \stepcounter{externalizedFigure}
    \fi
    \if\figureMode5
       \setcounter{externalizedSubFigure}{0}
       \stepcounter{externalizedFigure}
    \fi
    \input{#1}}

\newcommand{\includeSubsequentSubFigureOwn}[1]{
    \if\figureMode1
        \includegraphics[width=\textwidth]{example-image-a}
    \fi
    \if\figureMode2
        \input{#1}
    \fi
    \if\figureMode3
        \tikzexternalenable
        \stepcounter{externalizedSubFigure}
        \renewcommand{\externalizedFigurePath}{\filePrefix\theexternalizedFigure\alph{externalizedSubFigure}}
        \tikzsetnextfilename{\externalizedFigurePath}
        \input{#1}
        \tikzexternaldisable
    \fi
    \if\figureMode4
       \stepcounter{externalizedSubFigure}
       \includegraphics[scale=1]{\tikzpath\filePrefix\theexternalizedFigure\alph{externalizedSubFigure}.pdf}
    \fi
    \if\figureMode5
       \stepcounter{externalizedSubFigure}
       \scalebox{0}{
            \input{#1}
        }
       \includegraphics[scale=1]{\tikzpath\filePrefix\theexternalizedFigure\alph{externalizedSubFigure}}
    \fi
}

\makeglossaries

\pgfplotsset{compat=1.16}
\begin{document}
\begin{frontmatter}
\title{Modelling the response of a turbulent jet flame to acoustic forcing in a linearized framework using an active flame approach}

\author[TUB,myCorr]{Thomas L. Kaiser}
\ead{t.kaiser@tu-berlin.de}
\author[TUM]{Gregoire Varillon}
\author[TUM]{Wolfgang Polifke}
\author[KIT1]{Feichi Zhang}
\author[KIT1,KIT2]{Thorsten Zirwes}
\author[KIT1]{Henning Bockhorn}
\author[TUB]{Kilian Oberleithner}

\cortext[myCorr]{Corresponding author}

% TUB
\address[TUB]{
Laboratory for Flow Instabilities and Dynamics, 
Institute of Fluid Dynamics and Technical Acoustics, 
Technische Universit\"{a}t Berlin, 
M\"{u}ller-Breslau-Straße 8, 
10623 Berlin, Germany}
% TUM
\address[TUM]{
Technische Universit\"{a}t M\"{u}nchen, 
School of Engineering and Design,
Department of Engineering Physics and
Computation,
Boltzmann Str. 15, 
85747 Garching, Germany}
% KIT 1
\address[KIT1]{Engler-Bunte-Institute, 
Division of Combustion Technology, 
Karlsruhe Institute of Technology, 
Engler-Bunte-Ring 1, 
76131 Karlsruhe, Germany}
% KIT 2
\address[KIT2]{Steinbuch Centre for Computing (SCC), 
Karlsruhe Institute of Technology, 
Hermann-von-Helmholtz-Platz 1, 
76131 Karlsruhe, Germany}

\begin{abstract}
This study performs a linear mean field analysis of a turbulent reacting methane-air jet flame, with the goal of predicting the response of the reacting flow to upstream acoustic actuation. Unlike previous studies, this work develops and applies an active flame approach by taking the heat release oscillations of the flame resulting from the acoustic fluctuations into account. For an active flame approach in the linear mean field analysis, a linearized combustion model is necessary. Linearizing \gls{LES} and \gls{DNS} combustion models leads to closure problems, making their application in this context troublesome, whereas \gls{RANS} combustion models prove to circumvent this problem making them suitable candidates for this purpose. The \gls{RANS} combustion models are linearized around the temporal mean state variables of the turbulent jet flame, which is obtained by \gls{LES}. An a priori analysis shows that a linearized \gls{RANS}--\gls{EBU} model is the best suited among all investigated combustion models for the investigated set-up and reproduces with high accuracy the fluctuations in reaction rate obtained in the \gls{LES}. Furthermore, the linearized governing equations of the flow including the linearized \gls{EBU} model for the reaction rate are solved for incoming acoustic perturbations. The response modes show that the reaction rate oscillations are caused by Kelvin--Helmholtz vortex rings, which perturb the jet flame. The results are in good agreement with the \gls{LES} simulations in terms of the mode shapes of both reaction rate and velocity fluctuations. This study represents a basis for linear mean field analysis of turbulent flames and monolithic modelling approaches for thermoacoustic instabilities in gas turbine combustors.
\end{abstract}
\glsresetall

\begin{keyword}
\texttt{
linear modelling, flame modelling, turbulent flames, thermoacoustics}
\end{keyword}
\end{frontmatter}

\newpage
%\mbox{}
%\begin{multicols}{2}
%\printnomenclature
%\end{multicols}
%\begin{multicols}{2}
%\printglossaries
%\end{multicols}
\newpage

\section{Introduction}%%%%%%%%%%%%%%%%%%%%%%%%%%%%%%%%%
The dynamics of turbulent flames and the inherent fluctuations in heat release rate have been in the focus of gas turbine and rocket engineers for decades. First of all, the resulting fluctuations in heat release rate amplify noise emissions of gas turbines, becoming increasingly relevant in civil flight engines~\cite{Dowling2015}. Secondly, the heat release fluctuations can phase lock with acoustic chamber modes, leading to the self-reinforcing mechanism of a thermoacoustic instability~\cite{candel2002,lieuwen2005combustion,Poinsot2017}. These instabilities have various negative effects, such as increased NOx emissions, restriction of the operation range, or, in the worst case, damage of gas turbine components. Especially for combustion of hydrogen from renewable sources, these topics are assumed to pose an even more detrimental problem to the safe operation of gas turbines. In connection with the transition of the energy system, it is therefore of great interest to improve the understanding and modelling of flame dynamics. 

Flame dynamics and the therefrom resulting effects of combustion noise and thermoacoustic instabilities are multiphysics problems of significant complexity. Besides experimental investigations, \glspl{LES} have proven to be a valuable tool to study either phenomenon, allowing to take all relevant physical mechanisms into consideration. These methods yield valuable information for data driven approaches. Nevertheless, they often fail to provide sufficient physical insight into thermoacoustic instabilities and flame noise in order to find straightforward ways for their suppression or mitigation. Based solely on these tools, an optimization process quickly becomes an elaborate iterative cycle of trial and error. Especially with regards to thermoacoustic instabilities, reduced and low order methods provide relief. Examples in this context are network models (see e.g. Merk~\cite{Merk1957}, Dowling~\cite{Dowling1995}, Schuermans~\cite{schuermans2003modeling}) and Helmholtz solvers (see e.g. Silva et al.~\cite{silva2013Limit} and Mensah et al.~\cite{Mensah2016}). These methods, however, make use of considerable simplifications of the relevant physics of the turbulent reacting flow, and therefore, use strong assumptions, such as zero mean flow velocities, which may not be justified in real world applications. Furthermore, they describe the driving force of the thermoacoustic feedback cycle, i.e. the flame response resulting from acoustic perturbation, as a ``black box'', preventing the enlightenment of the precise mechanisms leading to the instability. A method with the potential to bridge these gaps, allowing the understanding and and control flame dynamics without drawing on the strong simplifications of today's reduced order models, would be the application of linearized mean field analysis to a turbulent reacting flow.

Linear mean field analysis, i.e. the analysis of the governing equations of the flow linearized around its temporal mean, has contributed significantly in recent years to the understanding of flow dynamics in both laminar and turbulent non-reacting flows. 
Barkley~\cite{Barkley2006} performed a hydrodynamic linear stability analysis around the mean flow of a cylinder in crossflow investigating the Kármán vortex street. He showed that the stability analysis yields accurate results for Reynolds numbers above the bifurcation point if it is performed around the temporal mean flow instead of the unperturbed  base flow. A type of linearized mean field analysis particularly contributing to the understanding of the generation of turbulence is the \gls{RA}. This type of analysis focusing on the elucidation of dominant amplification mechanisms in the flow, goes back to early works investigating the transition to turbulence in linearly stable flows (Trefethen et al.~\cite{Trefethen1993}, Butler and Farrell~\cite{Butler1992}, Farrell~\cite{Farrell1992}, Reddy et al.~\cite{Reddy1993,Reddy1993a}). Later, the method helped to deepen the understanding of turbulence in multiple configurations such as boundary layers~\cite{Cossu2009,Sipp2013}, jets flows~\cite{Garnaud2013a,beneddine2017,schmidt2018}, airfoil wakes~\cite{Abreu2018}, backward facing steps as well as channel~\cite{Abreu2020b,Morra2019a}, pipe~\cite{Abreu2020,Mckeon2010}, and Couette flows~\cite{Hwang2010}. Recent studies of Pickering et al.~\cite{pickering2020b, pickering2021}, applying the \gls{RA} to circular round jets at high Mach numbers to investigate the sound emissions of coherent turbulent structures are of particular interest to the case of flame noise.

In recent years, linear mean field analysis has been applied increasingly to turbulent flows in combustion systems. \tlk{One aspect which has been studied extensively in this context is the \gls{PVC} in swirling jet flows~\cite{Oberleithner2011}}. Juniper~\cite{juniper2012} and later Tammisola and Juniper~\cite{Tammisola2016} and Kaiser et al.~\cite{kaiser2018stability} used hydrodynamic linear mean field analysis to investigate the cause of a \gls{PVC} in the non-reacting flows of swirled real engine fuel injection systems. It was furthermore shown by Kaiser et al.~\cite{kaiser2020examining} that the information from this type of analysis can be used to adapt the fuel injector geometry in order to suppress the \gls{PVC}.

Besides the work on non-reacting flows, some studies addressed turbulent flames using the linear mean field analysis~\cite{Emerson2016,Frederick2018}. Oberleithner et al.~\cite{oberleithner2013Why} used a local linear stability analysis to investigate the effect of stratification of the temporally averaged density field neglecting density fluctuations. With this approach they provided an explanation why flames tend to suppress a \gls{PVC}. Oberleither et al.~\cite{oberleithner2016} used the same approach to explain the mechanism leading to the attenuation of the gain of a \gls{FDF} with increasing magnitude of the acoustic actuation. Manoharan and Hemchandra~\cite{manoharan2015Absolute} extended this approach by taking into account an energy equation in the low Mach number limit, allowing to address density fluctuations. In this way they investigated the interaction of fluctuations in the baroclinic torque and vorticity in the configuration of a turbulent flame anchored on a backward facing step. In analogy to a very similar approach addressing a thermoacoustic mode with a Helmholtz solver~\cite{silva2013Limit}, this approach was later termed ``passive flame'' approach by Kaiser et al.~\cite{kaiser2021Modeling}. They applied it in a global resolvent analysis to model the response of a turbulent swirl flame to acoustic perturbations. Using linear mean field analysis, Lueckhoff et al.~\cite{luckoff2021} and Frederick et al.~\cite{Frederick2018} showed that a mean field adaptation via a control of the \gls{PVC} can have a saturating effect on the symmetric vortex shedding in reacting jet flows. Most recently, Casel et al.~\cite{Casel2022} used the passive flame approach in a resolvent analysis in order to investigate the dominant coherent structures in a turbulent Bunsen flame.

All the above cited studies on linear mean flow analysis of turbulent reacting flows have in common to omit fluctuations of the heat release rate, which is a vital aspect of flame dynamics and of central importance to thermoacoustic instabilities and flame noise. In contrast, several studies included linearized combustion models termed here ``active flame'' approach, to study laminar flame dynamics. Early works, e.g. Pelce and Clavin~\cite{pelce1982}, used simple reaction schemes to investigate the Darrieus-Landau flame instability. More recently, Blachard et al.~\cite{blanchard2015} linearized a one-step Arrhenius reaction rate expression and used it to explain the interaction between hydrodynamic vortices and the laminar flame front. Albayrak et al.~\cite{AlbayBezgi17} as well as Avdonin et al.~\cite{avdonin2018} showed that the \gls{FTF} of laminar flames obtained with a one-step chemical reaction can be reproduced with the linearizing approach with high accuracy. Furthermore, Albayrak et al.~\cite{AlbayBezgi17}  identified in this framework an interaction mechanism between the radial velocity component of inertial waves and swirl flames, while  Avdonin et al.~\cite{avdonin2018} computed frequencies and growth rates of  thermoacoustically unstable  modes. Meindl et al.~\cite{meindl2021} showed that incorporating the flame dynamics via a global \gls{FTF} in the linearized Navier--Stokes equations causes spurious entropy waves, while the direct linearization of the flame dynamics (in their case a two step reaction mechanism) solved this problem. Recently, a \gls{RA} was applied to a laminar flame by Wang et al.~\cite{wang2022} in order to identify the amplification mechanisms leading to the flame dynamics, with specific focus on intrinsic thermoacoustic instabilities.

The above cited studies using an active laminar flame approach in the mean field analysis allowed to significantly deepen the understanding of the dynamics of laminar flames. Transferring this approach to  turbulent flames would open new pathways to investigate the highly relevant phenomena of thermoacoustics as well as direct and indirect combustion noise. A linear stability analysis using linear mean field analysis would allow to investigate thermoacoustic instabilities occurring in gas turbine combustors in a monolithic framework. This would provide insight into the entire feedback cycle, giving fundamentally new possibilities to study this phenomenon. The capability to predict the effect of geometry adaptations on an unstable mode, as conducted by Kaiser et al.~\cite{kaiser2020examining}, or even by the use of adjoint methods as discussed by Tammisola and Juniper~\cite{Tammisola2015} would be on the horizon. Addressing direct combustion noise, the \gls{RA} including an active flame approach would allow to directly tackle noise emissions in analogy to the work on non-reacting jets of Pickering et al.~\cite{pickering2020b,pickering2021}.

This study takes on the challenge of developing an active flame approach for a linear mean field analysis of a turbulent reacting flow. The configuration under investigation is a perfectly premixed, turbulent methane-air flame at an equivalence ratio of $\phi = 0.9$ and a Reynolds number of $Re \approx 8,000$. \gls{LES} provides the temporal mean flow, which is used as input to the linear mean field analysis. The \gls{LES} furthermore yields the response of the flame to upstream harmonic acoustic actuation. The goal of the present study is to develop a linearized combustion model, which reproduces the response of the turbulent flame to this acoustic actuation.

The following text is structured as follows: Section~\ref{ch:LES} discusses the \gls{LES} setup and  strategy and mean fields resulting from these computations. 
Subsequently, the linearized reacting flow equations for the turbulent flame under investigation will be discussed in Section~\ref{ch:LinearizationConservationEquations}. Section~\ref{ch:activeFlame} focuses on three candidate combustion models and their linearization. These three combustion models are evaluated in Section~\ref{ch:apriori}, by using an a priori analysis against \gls{LES} results. The best suited flame model based on the a priori analysis is applied in combination with the entire set of linearized governing equations in an a posteriori analysis in Section~\ref{ch:aposteriori}. Finally, Section~\ref{ch:conclusion} concludes the results of the present study.

\section{Large eddy simulations}\label{ch:LES}
\subsection{Numerical setup for \gls{LES}}
 
The configuration under investigation is a Bunsen-type, round jet-flame operated with a premixed methane/\-air mixture at an equivalence ratio $\Phi=0.9$ and $T_\text{u}=300$~K, $p_0=1$~atm. The diameter of the nozzle is $D=20$ mm and the bulk flow velocity is $5\, \text{m}\,\text{s}^{-1}$. The three-dimensional computational domain includes a part of the nozzle, with inlet of the fresh gas located at 1$D$ upstream the nozzle exit. The nozzle is connected to a large cylindrical chamber with a length of 25$D$ and a diameter of 8$D$. The computational grid consists of 3.9 million hexahedral cells and is locally refined along the shear layer of the jet (see left image of Fig.~\ref{fig:LES}), with the smallest grid resolution of 0.12 mm in the radial direction. The Reynolds number based on the nozzle diameter and the bulk velocity at the exit plane is $Re\approx 8,000$.

The open-source code OpenFOAM has been used to perform the simulations, where discretizations of convective and diffusive terms are based on an unbounded, central difference scheme of 2nd order accuracy. The Smagorinsky sub-grid-scale model is applied in combination with a \gls{TFC} combustion model (see Section~\ref{:ch:nonLinearCombustionModelling}).
No-slip and adiabatic conditions have been applied for the nozzle and chamber walls. At the outlet, gradients of all flow variables and chemical scalars were set to zero. A turbulence inflow generator~\cite{galeazzo2021,klein2003} has been used to provide spatially and temporally correlated velocity fluctuations with pre-defined turbulence parameters at the inlet. The partially non-reflecting boundary condition proposed by Poinsot and Lele~\cite{nrbc1992} has been applied to the inlet and outlet boundaries to avoid unphysical reflection of acoustic waves. The time step was set to 2.5$\cdot 10^{-5}$ s, leading to a Courant-Friedrichs-Lewy (CFL) number of approximately 0.7. 

%%The numerical setup is based on the Kobayashi flame~\cite{Kobayashi2002} at atmospheric conditions and was validated against the experimental results by Zhang et al.~\cite{zhang2020}. For the sake of simplifying the linearization process, two changes were made with respect to the experimental set-up: Firstly, adiabatic walls allow to neglect heat losses to walls. Secondly, a prolongation of the combustion chamber allows to investigate coherent structures with very large wave lengths at low frequencies.

The numerical setup is based on a canonical configuration with a Bunsen-type, round-jet flame, which is specifically designed to study the flame response to acoustic forcing. For the sake of simplifying the linearization process, the flame is confined by adiabatic walls, which allows to neglect heat losses to walls. In addition, a relatively long combustion chamber is used to investigate coherent structures with very large wave lengths at low frequencies. 

\subsection{Modelling strategy of turbulent combustion in \gls{LES} }\label{:ch:nonLinearCombustionModelling}
Reaction progress is modelled by an additional transport equation of a reaction progress variable, $c^\text{F}$, where the superscript F denotes a Favre average. To evaluate the chemical scalars based on $c$, a 1D freely-propagating, unstrained premixed flame has been pre-computed with the thermo-chemical library Cantera, using the GRI-3.0 reaction mechanism~\cite{gri}. The progress variable $c$ is then evaluated from the chemically bound oxygen and the species mass fractions $Y_k$ have been projected onto the $c$-space, leading to a chemistry look-up table $Y_k = f(c)$. 
The \gls{TFC} model~\cite{zhang2019} has been used to account for the flame-turbulence interaction:
\begin{equation}
   \overline{ \dot{\Omega}} = \rho_\text{u} \frac{s_\text{t}^2}{D_\text{t}+D} c^\text{F}\left(1-c^\text{F}\right).
\end{equation}
Here $\overline{\dot{\Omega}}$ is the reaction rate, $\rho_\text{u}$ the fresh gas density, $c^\text{F}$ the Favre averaged progress variable, while $D_\text{t}$ and $D$ are the turbulent sub-grid-scale and molecular diffusion of the progress variable, respectively. \gls{SGS}-turbulence flame interaction is modeled via a turbulent flame speed which is given by
\begin{equation}
    \frac{s_\text{t}}{s_\text{l}}= 1 + \frac{u_\text{t}}{s_\text{l}} \left( 1+ Da^{-2}\right)^{-0.25},
\end{equation}
where $s_\text{l}$ is the unstretched laminar flame speed and $u_\text{t}$ the turbulent \gls{SGS} intensity. The Damköhler number is modeled as $Da=\left( L_\text{sgs} / u_\text{sgs} \right)/ \left(\alpha_\text{u} / s_\text{l}^2\right)$, where $\alpha_\text{u}$ is the thermal diffusivity of the unburnt mixture. The turbulent intensity is approximated from the turbulent kinetic energy, $k$, by the relation $u_\text{t}=k^{0.5}$, while the \gls{SGS} length scale, $L_\text{sgs}$, corresponds to one tenth of the \gls{LES} filter width, $L_\text{sgs}=0.1\Delta$.

The \gls{TFC}-LES approach has been extensively applied to different flame configurations and has been discussed in comparison to other  models in e.g.~\cite{zhang2009,zhang2012,zhang2013,zhang2019,zhang2020}. Although \gls{TFC}-models originated from \gls{RANS}-simulations, these models are often also applied in \gls{LES}~\cite{Poinsot}. These LES-TCF models have been shown to perform similarly well when compared to models based on Arrhenius-like reaction rates on grids with coarse and medium resolution of the flame sheet.

\subsection{Temporal mean fields from \gls{LES}}
\begin{figure}[t]
    \centering
\input{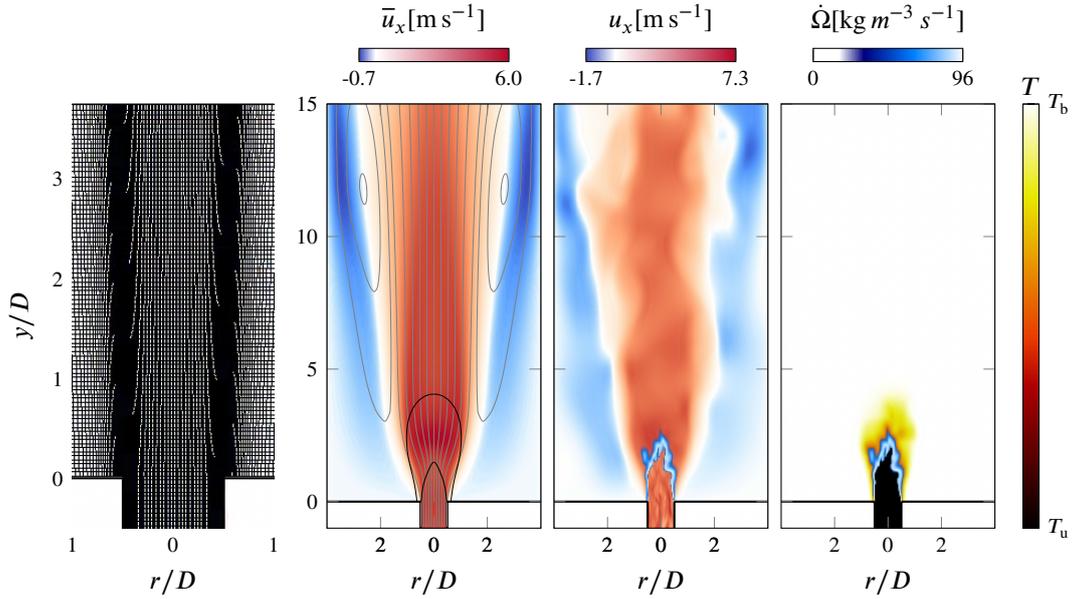}
    \caption{\gls{LES} setup and results; left: computational grid, focused on the flame region; center left: temporally averaged streamwise velocity field with streamlines (gray) and temporally averaged flame position indicated by the iso-contours, $\overline{c}=0.02$ and $\overline{c}=0.98$ (solid black lines); center right: instantaneous velocity field and flame position indicated by the instantaneous reaction rate; right: instantaneous temperature field and flame position indicated by the instantaneous reaction rate}
    \label{fig:LES}
\end{figure}

Some results of the applied \gls{LES} are illustrated in Fig.~\ref{fig:LES}. The center left column illustrates the streamwise velocity component in combination with streamlines (gray). Furthermore, the iso-contours of $\overline{c}=0.02$ and $\overline{c}=0.98$ (black lines) show the position of the temporal average of the flame. In addition to the acoustically unperturbed simulations, \gls{LES} with acoustic forcing from upstream direction are conducted. Three forcing frequencies are investigated, $f=25\, \text{Hz}$, $f=50\, \text{Hz}$, and $f=75\, \text{Hz}$. The acoustic actuation velocity is 15\,\% of the bulk speed. The results of these forced simulations will be discussed together with the results of the linear mean flow analysis in Sections~\ref{ch:apriori} and \ref{ch:aposteriori}. The center-right illustration depicts an instantaneous field of streamwise velocity, while the right hand illustration shows the corresponding instantaneous temperature field. The latter two plots are overlaid by the instantaneous reaction rate in the blue-to-white color map. Note that due to the adiabatic walls, the burnt gas ($c=1$)  temperatures are close to the adiabatic flame temperature, $T_\text{b}=2138 \, \text{K}$ in the entire region downstream of the flame. As a consequence, the temperature can be estimated as a function of the progress variable. This simplifies deriving the linearized equations, as will be discussed in the following section.

\section{Linearized turbulent reacting flow equations}\label{ch:linearizedFramework}
\subsection{Non-linear governing equations}
The flame configuration investigated in this work allows to draw on some simplifications, which drastically reduce the complexity of the set of equations subjected to linearization. These are:
\begin{enumerate}
    \item The flow can be assumed to be incompressible. This is because the flow speed is at low Mach numbers and the relevant acoustic wave lengths for the frequencies under investigation are significantly larger than the flame length.
    \item Heat losses to walls can be neglected, since in the \gls{LES} adiabatic \glspl{BC} are considered. In the burnt state, the temperature hence reaches the adiabatic flame temperature in the entire domain. As a consequence the temperature becomes a function of the progress variable alone.
\end{enumerate}
Besides these simplifications, we will assume constant heat capacities and specific gas constants. Based on these simplifications and assumptions, the temperature is only a function of the progress variable, resulting in
\begin{equation}
    {T} = T_\text{u} + \left( T_\text{b} - T_\text{u} \right) {c}, \label{eq:TOfC:nonlinear}
\end{equation}
where $T_\text{u} = 300\, \text{K}$ and $T_\text{b} = 2138 \text{K}$ are the unburnt and adiabatic flame temperature, respectively. Furthermore, at low Mach number conditions, the pressure in the ideal gas equation can assumed to be constant leading to
\begin{equation}
    \rho = \frac{\overline{p}}{RT} \approx  \frac{{p_0}}{RT}, \label{eq:perfectGas:nonlinear}
\end{equation}
where $R$ is the specific gas constant, which for methane-air combustion is approximately equal for the fully unburnt and burnt state. 
Eqs.~(\ref{eq:TOfC:nonlinear}) and (\ref{eq:perfectGas:nonlinear}) relate the progress variable and the fluid density, obviating the need for an energy balance equation. This reduces the complexity of the set of equations, which then consist of the balance equation of momentum, mass and a transport equation for the progress variable, reading 
\begin{subequations}
\begin{equation}
    \rho \left( \frac{\partial \mathbf{u}}{\partial t} 
    + \left( \mathbf{u} \cdot \nabla \right) \mathbf{u} \right)
    = - \nabla p 
    + \nabla \cdot \mu \left( 
        \nabla \mathbf{u} 
        + \left(\nabla \mathbf{u}\right)^\text{T} 
        - \frac{2}{3} \left( \nabla \cdot \mathbf{u} \right) \mathbf{I} 
    \right),
\end{equation}
\begin{equation}
\frac{\partial  {\rho}}{\partial t}
+ \nabla \cdot \left( \rho \mathbf{u} \right)
=0,
\end{equation}
\begin{equation}
    \rho \frac{\partial c}{\partial t} 
    + \rho \mathbf{u} \cdot \nabla c
    =\nabla \cdot D \nabla c 
    + \dot{\Omega}.
\end{equation}\label{eq:Governing:nonlinear}
\end{subequations}

\subsection{Linearization of the governing equations}\label{ch:LinearizationConservationEquations}
To linearize the set of Eqs.~(\ref{eq:TOfC:nonlinear})--(\ref{eq:Governing:nonlinear}) around a temporal mean state, the triple decomposition as suggested by Hussain and Reynolds~\cite{hussain1970themechanics,Reynolds1972} is applied to the state variables. It decomposes the state variables into a temporal mean, a coherently fluctuating part and a turbulent fluctuating part which is uncorrelated with the coherent fluctuation
\begin{equation}
     [\mathbf{u},p,\rho,T,c]
    = [\overline{\mathbf{u}},\overline{p},\overline{\rho},\overline{T},\overline{c}]
    + [\widetilde{\mathbf{u}},\widetilde{p},\widetilde{\rho},\widetilde{T},\widetilde{c}]
    + [\mathbf{u}^\prime,p^\prime,\rho^\prime,T^\prime,c^\prime].\label{eq:tripple}
\end{equation}
The definition of the coherently fluctuating part is given by the phase average at the frequency of the coherent fluctuation, indicated by the angled brackets:
\begin{equation}
    [\widetilde{\mathbf{u}},\widetilde{p},\widetilde{\rho},\widetilde{T},\widetilde{c}] 
    =\langle[\mathbf{u},p,\rho,T,c]  \rangle - [\overline{\mathbf{u}},\overline{p},\overline{\rho},\overline{T},\overline{c}].\label{eq:phaseAverage}
\end{equation}
The definition of the stochastic fluctuations follows then from Eqs.~(\ref{eq:tripple}) and (\ref{eq:phaseAverage}). Inserting this relation into the non-linear set of equations~(\ref{eq:TOfC:nonlinear})--(\ref{eq:Governing:nonlinear}), taking the difference between the phase average and the temporal average of the resulting equation and neglecting all terms, which are non-linear in coherent fluctuations, leads to the linearized set of equations for the coherent fluctuations in the time domain:
\begin{subequations}
\begin{equation}
    \overline{\rho} \frac{\partial \widetilde{\mathbf{u}}}{\partial t} 
    + \overline{\rho} \left( \left( \widetilde{\mathbf{u}} \cdot \nabla \right) \overline{\mathbf{u}} + \left( \overline{\mathbf{u}} \cdot \nabla \right) \widetilde{\mathbf{u}} \right)
    + \widetilde{\rho}  \left( \overline{\mathbf{u}} \cdot \nabla \right) \overline{\mathbf{u}}  
    + \reallywidetilde{{\rho  \left( {\mathbf{u}^\prime} \cdot \nabla \right) {\mathbf{u}^\prime}}}
    = - \nabla  \widetilde{p} 
    + \nabla \cdot {\mu} \left( 
        \nabla  \widetilde{\mathbf{u}} 
        + \left(\nabla  \widetilde{\mathbf{u}}\right)^\text{T} 
        - \frac{2}{3} \left( \nabla \cdot  \widetilde{\mathbf{u}} \right) \mathbf{I} 
    \right)\label{eq:momentum:linear:timedomain}
\end{equation}
\begin{equation}
    \frac{\partial  \widetilde{\rho}}{\partial t}
+ \nabla \cdot \left( \widetilde{\rho} \overline{\mathbf{u}} \right)
+ \nabla \cdot \left( \overline{\rho} \widetilde{\mathbf{u}} \right)
=0,\label{eq:mass:linear:timedomain}
\end{equation}
\begin{equation}
    \overline{\rho} \frac{\partial \widetilde{c}}{\partial t} 
    + \widetilde{\rho} \overline{\mathbf{u}} \cdot \nabla \overline{c}
    + \overline{\rho} \widetilde{\mathbf{u}} \cdot \nabla \overline{c}
    + \overline{\rho} \overline{\mathbf{u}} \cdot \nabla \widetilde{c}
    + \reallywidetilde{\rho {\mathbf{u}^\prime} \cdot \nabla {c}^\prime}
    =\nabla \cdot {D} \nabla \widetilde{c} 
    + \langle {\dot{\Omega}}\rangle
    -\overline{\dot{\Omega}},\label{eq:progressVariable:linear:timedomain}
\end{equation}
\end{subequations}
Equations~(\ref{eq:mass:linear:timedomain})--(\ref{eq:progressVariable:linear:timedomain}) correspond to the linearized balance equations of momentum, mass and the linearized transport equation for the progress variable. 

The fifth terms in the linearized equations of momentum and progress variable arise from the non-linear terms and are of the same order as the linear  perturbations~\cite{Reynolds1972}. It was shown by Viola et al.~\cite{Viola2014} for the momentum equation and Kaiser et al.~\cite{kaiser2021Modeling} for the balance equation of species that these can be modeled by turbulent contributions to viscosity, 
$\mu_\text{t}$, and species diffusivity, $D_\text{t}$, respectively. In this study, these are obtained using a \gls{TKE} model based on the mixing length:
\begin{equation}
    \mu_\text{t} = \overline{\rho} \nu_\text{t} = \overline{\rho} \beta l k^{0.5},\quad D_\text{t} = \frac{\mu_\text{t}}{\overline{\rho} \text{Sc}_\text{t}},\label{eq:eddyViscosity}
\end{equation}
where $\beta$ is a model parameter and set to $0.05$, $\text{Sc}_\text{t}$ is the turbulent Schmidt number, which is set to unity in this study, and $k$ is the turbulent kinetic energy, which is extracted from the \gls{LES} statistics. The parameter $l=l(x)$ is the shear layer thickness which is determined at every streamwise position in radial direction and therefore is a function of the streamwise coordinate only.

Finally, a linearized equation of state is needed to relate the coherent fluctuations of progress variable and density. This is obtained by linearizing Eqs.~(\ref{eq:TOfC:nonlinear}) and~(\ref{eq:perfectGas:nonlinear}), yielding
\begin{equation}
    \widetilde{\rho} = - \frac{\overline{\rho}}{\overline{T}} \widetilde{T}, \quad 
    \widetilde{T}= \left(T_\text{b}-T_\text{u}\right) \widetilde{c}.\label{eq:rhoOfCLin}
\end{equation}
Next, a harmonic ansatz of the form
\begin{equation}
    [\widetilde{\mathbf{u}}, \widetilde{p},\widetilde{\rho},\widetilde{c}] = [\widehat{\mathbf{u}}, \widehat{p},\widehat{\rho},\widehat{c}]\, 
    \text{exp} \left( - i \omega t\right) 
    + \text{c.c.},
\end{equation}
is introduced, transforming the linearized equations into the frequency domain. This leads to the following set of linear governing equations, which are addressed in this study:
\begin{subequations}
\begin{equation}
    - \text{i} \omega \overline{\rho}  \widehat{\mathbf{u}}
    + \overline{\rho} \left( \left( \widehat{\mathbf{u}} \cdot \nabla \right) \overline{\mathbf{u}} + \left( \overline{\mathbf{u}} \cdot \nabla \right) \widehat{\mathbf{u}} \right)
    + \widehat{\rho}  \left( \overline{\mathbf{u}} \cdot \nabla \right) \overline{\mathbf{u}}  
    = - \nabla  \widehat{p} 
    + \nabla \cdot \left( {\mu} + {\mu}_\text{t}\right)  \left( 
        \nabla  \widehat{\mathbf{u}} 
        + \left(\nabla  \widehat{\mathbf{u}}\right)^\text{T} 
        - \frac{2}{3} \left( \nabla \cdot  \widehat{\mathbf{u}} \right) \mathbf{I} 
    \right)\label{eq:momentum:linear:freqdomain}
\end{equation}
\begin{equation}
   - \text{i} \omega  \widehat{\rho}
+ \nabla \cdot \left( \widehat{\rho} \overline{\mathbf{u}} \right)
+ \nabla \cdot \left( \overline{\rho} \widehat{\mathbf{u}} \right)
=0,\label{eq:mass:linear:freqdomain}
\end{equation}
\begin{equation}
    - \text{i} \omega \overline{\rho} \widehat{c}
    + \widehat{\rho} \overline{\mathbf{u}} \cdot \nabla \overline{c}
    + \overline{\rho} \widehat{\mathbf{u}} \cdot \nabla \overline{c}
    + \overline{\rho} \overline{\mathbf{u}} \cdot \nabla \widehat{c}
    =\nabla \cdot \left( {D} + {D}_\text{t}  \right)\nabla \widehat{c} 
    + \widehat{\dot{\Omega}},\label{eq:progressVariable:linear:freqdomain}
\end{equation}
\begin{equation}
    \widehat{\rho} = - \frac{\overline{\rho}}{\overline{T}} \widehat{T}, \quad 
    \widehat{T}= \left(T_\text{b}-T_\text{u}\right) \widehat{c}.
\end{equation}\label{eq:linear:freqdomain}
\end{subequations}
The coherent fluctuation in reaction rate, $\widehat{\dot{\Omega}}_c$ appearing in Eq.~\ref{eq:progressVariable:linear:freqdomain} is neglected in the passive flame approach. The central goal of this study, however, is to develop and apply an active flame approach by taking this term into account. In the following section possibilities of finding an expression for this quantity are explicated.

\section{Active flame approach}\label{ch:activeFlame}
An active flame approach takes  the heat release oscillations resulting from a coherent perturbation of the state variables around the temporal mean state into account. In its most general form, an active flame approach is given by the relation
\begin{equation}
\widehat{\dot{\Omega}} = f(\overline{\Phi},\widehat{\Phi}),\label{eq:ActiveFlame}
\end{equation}
with the state variables, $\Phi$, in this study given by $\Phi=[\mathbf{u},p,c]$. This section discusses, how an expression for $\widehat{\dot{\Omega}}$ can be derived, and which non-linear combustion models qualify for linearization.

\subsection{Mean reaction rates from linearization of combustion models}
Various combustion models have been developed to tackle different flame regimes and numerical approaches. To examine their suitability to be applied in the linearized mean field analysis, combustion models can be usefully categorized with respect to the belonging to the different groups of \gls{CFD} simulation approaches. The first category in this context is the group of combustion models used in \gls{DNS} and \gls{LES}. They yield in the case of \gls{DNS} an instantaneous reaction rate, $\dot{\Omega}$ or in the case of \gls{LES} a control volume averaged rate as a function of the instantaneous or  sub grid filtered state variables $\Phi$, respectively. These  combustion models can be written in their most general form as
\begin{equation}
    \dot{\Omega}_\text{DNS/LES} = f(\Phi)\label{eq:general_omega}
\end{equation}
Introducing the triple decomposition in Eq.~\ref{eq:general_omega} with subsequent phase and temporal averaging as performed in Section~\ref{ch:linearizedFramework} leads to numerous unclosed terms, which have to be modelled. An alternative to this approach neglects turbulent fluctuations, $\Phi^\prime$, by expanding the variables into a Taylor series around the temporal mean and cutting the series after the linear term. The effect of neglecting the turbulent fluctuations will be discussed in hindsight. 
The Taylor expansion then reads
\begin{equation}
    {\dot{\Omega}} 
    \approx f (\overline{\Phi})
    +\frac{\partial f}{\partial \Phi}\bigg\rvert_{\Phi=\overline{\Phi}} \widehat{\Phi}.\label{eq:Taylor:DNS}
\end{equation}
A temporal average and a phase average of the linear Taylor expansion yield
\begin{equation}
     \overline{\dot{\Omega}} \approx \Omega ( \overline{\Phi} ),\label{eq:TaylorTemp}
\end{equation}
and an expression for the coherent fluctuation in reaction rate
\begin{equation}
     \widehat{\dot{\Omega}} \approx \frac{\partial f}{\partial \Phi}\bigg\rvert_{\Phi=\overline{\Phi}} \widehat{\Phi},\label{eq:TaylorPhase}
\end{equation}
respectively. These relations hold well for the linearization around the state variables of a stationary laminar flame, and lead to linearized reaction models which were used for example by Avdonin et al.~\cite{avdonin2018}, Wang et al.~\cite{wang2022} and Meindl et al.~\cite{meindl2021}. For turbulent flames however, it is a well known problem from \gls{RANS} approaches~\cite{Poinsot} that the relation given by Eq.~(\ref{eq:TaylorTemp}) does not hold. This is because neglecting the influence of stochastic fluctuations on reaction rates consisting of strongly non-linear functions of state variables and the closure problems arising from that. In the presence of turbulence, an equally severe closure problem occurs also for the linearized reaction rate given by Eq.~(\ref{eq:TaylorPhase}). As a consequence, it cannot be expected that this description of the linearized reaction rate is accurate without addressing the arising closure problem explicitly. The analogous closure problem exists in the \gls{RANS} framework. However, here a different approach has prevailed to circumvent the above closure problem and the turbulence-flame interaction is modelled using turbulent time scales (as done in \gls{EBU} models) or a turbulent flame speed closure is applied to model this effect (\gls{TFC} models). These kind of models directly yield the temporal average of the reaction rate as a function of the temporal average of the state variables. In their most general form, they can be written as
\begin{equation}
    \overline{\dot{\Omega}}_\text{RANS} = f(\overline{\Phi}).
\end{equation}
An approach similar to turbulent flame modelling in \gls{RANS} simulations will be performed in this study to treat this closure problem in the linearized mean field analysis. A Taylor series expansion taking only the linear terms yields the reaction rate resulting from the temporal mean state variables, which are perturbed by a small coherent fluctuation:
\begin{equation}
    {\dot{\Omega}}
    \approx f (\overline{\Phi})
    +\frac{\partial f}{\partial \overline{\Phi}} \widehat{\Phi}. \label{eq:Taylor:RANS}
\end{equation}
A temporal average of the equation gives again the relation in Eq.~(\ref{eq:TaylorTemp}), while a phase average yields the coherent fluctuation in reaction rate
\begin{equation}
    \widehat{\dot{\Omega}}
    \approx 
    \frac{\partial f}{\partial \overline{\Phi}} \widehat{\Phi}. \label{eq:Taylor:RANS:Phase}
\end{equation}
The difference to the previously discussed \gls{DNS} and \gls{LES} models is that the \gls{RANS} combustion models fulfill Eq.~(\ref{eq:TaylorTemp}) by definition due to the absence of a closure problem. Analogously, no additional turbulence closure problem occurs in Eq.~(\ref{eq:Taylor:RANS:Phase}), which qualifies this kind of models to be applied in a linearizing approach. Using \gls{RANS} combustion  models may raise an inconsistency, because  \gls{RANS} simulations are known to yield non-accurate descriptions of the temporal mean fields of reacting flows in most configurations. If the temporal mean flow, however, is not accurate it cannot be expected that a linearization around that mean variables yields accurate results. A solution to this inconsistency is to take the temporal mean variables of the flow obtained by higher fidelity approaches, such as \gls{LES} or \gls{DNS}. This approach is pursuit in this work. The corresponding data from \gls{LES} is illustrated in Section \ref{ch:LES}.

Following this approach, it cannot be expected a priori from a linearized combustion model, however, to yield a correct description of the coherent fluctuations in reaction rates if the corresponding non-linear combustion model does not reproduce the temporal mean reaction rate. Therefore, it is necessary that \gls{RANS} combustion  models at least reproduce the mean reaction rates from \gls{LES} according to
\begin{equation}
     \overline{\dot{\Omega}}_\text{DNS,LES} \approx \dot{\Omega}_\text{RANS} ( \overline{\Phi}),\label{eq:consistency}
\end{equation}
before being applied in the linearized mean field analysis. This aspect will be discussed in the following section for three combustion models.

\subsection{Non-linear combustion models and their linearization}\label{ch:linReactionRate}

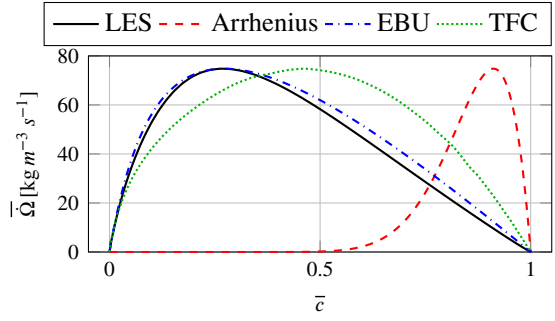
\begin{figure} 
    \centering
{\pgfplotsset{
    compat=1.3, 
    every axis/.append style={scale only axis,% axis on top,
    height=2.6cm, width=0.4\linewidth, xmin=-0.05, xmax=1.05,ymin=0,ymax=80,label style={font=\footnotesize},tick label style={font=\footnotesize}
    }
}
    \pgfplotsset{scaled x ticks=false}
\begin{tikzpicture}[scale=1.0000]

\begin{axis}[
scale only axis,
grid=both,
name=real,
ylabel style={align=center},
ylabel={$\overline{\dot{\Omega}}\,[\text{kg}\,m^{-3}\,s^{-1}]$},
xlabel=$\overline{c}$,
legend pos=outer north east,
legend cell align={left},
xtick={0,0.5,1},
legend style={
    at={(1.0,1.05)},
    anchor=south east,
    legend columns=-1,
    }
    ]
\addplot [color=black,solid,smooth,  thick] table [x = c , y=rr_LES, col sep=comma, solid,color=black]{Data/reactionRates.csv};\addlegendentry{LES};
\addplot [color=red,dashed,smooth,  thick] table [x = c , y=rr_Ar, col sep=comma, solid,color=black]{Data/reactionRates.csv};\addlegendentry{Arrhenius};
\addplot [color=blue,dashdotted,smooth,  thick] table [x = c , y=rr_EBU, col sep=comma, solid,color=black]{Data/reactionRates.csv};\addlegendentry{EBU};
\addplot [color=green!70!black,densely dotted,smooth,  thick] table [x = c , y=rr_TFC, col sep=comma,color=black]{Data/reactionRatesTFC.csv};\addlegendentry{TFC};
\end{axis}

\end{tikzpicture}
}
    \caption{Non-linear reaction rates of the Arrhenius model (Eq.~(\ref{eq:Arrhenius:nonLinear})), the \gls{EBU} model (Eq.~(\ref{eq:EBU:nonlinear})), and the \gls{TFC} model (Eq.~(\ref{eq:TFC:nonlinear}))}
    \label{fig:reactionRates:nonlinear}
\end{figure}

In this section three different non-linear combustion models are evaluated with respect their ability to reproduce the mean reaction rates and hence constitute valid candidates for the linearization. The evaluation will be performed against the temporally averaged fields provided by the \gls{LES} according to Eq.~(\ref{eq:consistency}). For this, the temporally averaged reaction rate as a function of the temporal average of the progress variable extracted from the \gls{LES} along the jet center line of the flame displayed in Fig.~\ref{fig:LES} is shown in Fig.~\ref{fig:reactionRates:nonlinear} (black solid line) and compared with that from the three combustion models.

The first example is an Arrhenius type reaction rate model.  Although this model is empirical, it describes chemical kinetics of single reactions with high accuracy, and is, therefore, mostly applied in \gls{DNS} and -- mostly in combination with a thickened flame model -- in \gls{LES}. Using a progress variable and the assumptions drawn for the linearized mean field analysis in Section~\ref{ch:LinearizationConservationEquations}, an Arrhenius type source term can be formulated as
\begin{equation}
    \overline{\dot{\Omega}}_\text{Ar} = A_\text{Ar} \overline{\rho} \left(1-\overline{c}\right) \text{exp}\left( -\frac{T_\text{a}}{\overline{T}}\right), \label{eq:Arrhenius:nonLinear}
\end{equation}
where $A_\text{Ar}$ is the Arrhenius pre-factor, and $T_\text{a}$ the activation temperature, which is set here to  $T_\text{a}=26,000\, \text{K}$. Note that both the temperature and the density can be related to the progress variable using Eqs.~(\ref{eq:TOfC:nonlinear}) and (\ref{eq:perfectGas:nonlinear}), such that Eq.~(\ref{eq:Arrhenius:nonLinear}) can be written as a function of $\overline{c}$ alone. Furthermore, the Arrhenius model is formulated as a \gls{RANS} type model, since, as discussed in the previous section, \gls{LES} and \gls{DNS} models would lead to significant closure problems. In this form, the model is only applicable to few drastically simplified configurations. Nevertheless, this model will be discussed in this work for illustrative purposes. For the sake of comparing the model to the \gls{LES} results, the prefactor is set to a constant value, causing the identical maximum of the temporally averaged reaction rates in the \gls{LES} and the Arrhenius model. The resulting temporal mean of the reaction rate is given by the dashed red line in Fig.~\ref{fig:reactionRates:nonlinear}. As expected, the Arrhenius reaction rate based on the mean variables is not capable of reproducing the turbulent reaction rate profile as obtained by the \gls{LES}, due to disregarding the arising closure problem. While the temporal mean of the reaction rate based on the \gls{LES} peaks at approximately $\overline{c}\approx0.28$, the highest value in $\overline{\dot{\Omega}}$ in the Arrhenius model is obtained at $\overline{c}\approx 0.95$. Furthermore, the Arrhenius model shows, in contrast to the \gls{LES} results, negligible reaction rate for values of the progress variables below $\overline{c}=0.5$. As a consequence, the Arrhenius model cannot be expected to provide accurate results if applied to an active flame approach in a linearized mean field analysis.

The second combustion model is the \gls{TFC} model and is of \gls{RANS} type including the magnitude of the gradient of the temporally averaged progress variable:
\begin{equation}
    \overline{\dot{\Omega}}_\text{TFC} = \rho_\text{u} s_\text{t} |\nabla \overline{c}| 
    = A_\text{TFC} |\nabla \overline{c}| \label{eq:TFC:nonlinear}
\end{equation}
In Eq.~\ref{eq:TFC:nonlinear} $\rho_\text{u}$ is the density of the unburnt mixture and $s_\text{t}$ the turbulent flame speed taking into account the turbulence-flame interactions. Assuming that the turbulent flame speed does not vary significantly along the jet center line, these two quantities can be lumped together into a constant prefactor, $A_\text{TFC}$. Like in the Arrhenius model, for the sake of comparison against the \gls{LES} results, this prefactor is set here to a constant value such that the maxima of the temporal average of reaction rates coincides with the one in the \gls{LES}. The resulting reaction rate is plotted by the green dotted line in Fig.~\ref{fig:reactionRates:nonlinear}. Although the model gives a significantly better reproduction of the turbulent reaction rate profile from the \gls{LES} in comparison to the Arrhenius model, predominant differences are visible. The \gls{TFC} reaction rate peaks at a value of $\overline{c}\approx0.47$, which is significantly higher than the one seen in the \gls{LES}.

The third and final combustion model under investigation is of \gls{EBU} type and reads
\begin{equation}
    \overline{\dot{\Omega}}_\text{EBU} = C_\text{EBU} \frac{\epsilon}{k} \overline{\rho}\, \overline{c} \left(1-\overline{c}\right)
    = A_\text{EBU} \overline{\rho}\, \overline{c} \left(1-\overline{c}\right).\label{eq:EBU:nonlinear}
\end{equation}
Here $C_\text{EBU}$ is a model parameter, and $\epsilon$ the turbulent dissipation rate. In a \gls{RANS} framework, turbulence-flame interactions are taken into account by a turbulent time scale given by the ratio $\frac{\epsilon}{k}$. As a consequence, typically a $k$-$\epsilon$-turbulence model is used in combination with this approach~\cite{Poinsot}. In analogy with the previous combustion models, we lump the quantities modelling the turbulence-flame interactions into a constant prefactor, such that the maximum of the reaction rate coincides with the one obtained from the \gls{LES}. The result is illustrated by the blue dash-dotted line in Fig.~\ref{fig:reactionRates:nonlinear}. It appears that the \gls{EBU}-model shows very good agreement with the \gls{LES} mean reaction rate for all $\overline{c}$.

Summing up, it appears that among the investigated combustion models the \gls{EBU} model reproduces best the mean reaction rates from \gls{LES} and is, therefore, suited for application in an active flame approach in the mean field analysis based on the mean fields provided from the \gls{LES} applied in this study. The \gls{EBU} model is followed by the \gls{TFC} model and finally the Arrhenius model. This will be investigated and discussed in detail in the Section~\ref{ch:apriori}.

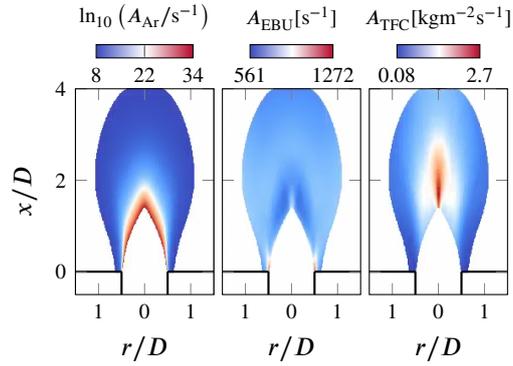
\begin{figure}
    \centering
{\pgfplotsset
{
    compat=1.3, 
    every axis/.append style=
    {
        scale only axis,
        % define the custom colormap
        colormap={my colormap}{
                rgb255=(59, 76, 192),
                rgb255=(255, 255, 255  ),
                rgb255=(180, 4, 38),
        },
        title style={
            yshift=-4.0pt,
            font=\footnotesize,
                    },
    },
    xlabel=$r/D$,
    xtick={-1,0,1},
    xticklabels={1,0,1},
}

\begin{tikzpicture}[scale=1.0]
    \node (reference) {};
        \foreach \imodel/\model in {0/Ar,1/EBU,2/TFC}
        {
        
            \begin{axis}
                        [
                            at={($(reference.south west)+(\imodel*55pt,0)$)},
                            anchor=west, 
                            enlargelimits=false,
                            axis equal image,
                            name=last,
                            width=90pt,
                            xtick = {-1,0,1},
                            xlabel={},
                            ylabel={},
                            xticklabels={,},
                            yticklabels={,},
                            xmin=-1.5,xmax=1.5,ymin=-0.5,ymax=4,
                        ]
                        \addplot graphics [xmin=-1.5,xmax=1.5,ymin=-0.5,ymax=4] {pngs/prefactors/A_\model.png};
                        \addplot [color=black, mark=none,thick] coordinates { (-1.5,0) (-0.5,0)};
                        \addplot [color=black, mark=none,thick] coordinates { (-0.5,0) (-0.5,-0.5)};
                        \addplot [color=black, mark=none,thick] coordinates { (1.5,0) (0.5,0)};
                        \addplot [color=black, mark=none,thick] coordinates { (0.5,0) (0.5,-0.5)};
            \end{axis}
            \ifthenelse{\imodel=0}
        {
         \pgfplotsset{
                    colorbar horizontal,
                    ylabel={$x/D$},
                    point meta min=0,
                    point meta max=1,
                    colorbar style=
                    {
                        at={(0.5,1.15)},
                        anchor=south,
                        width=0.7*\pgfkeysvalueof{/pgfplots/parent axis width},
                        height={5pt},
                        title=$\text{ln}_{10}\left(A_\text{Ar}/\text{s}^{-1}\right)$,
                        xtick={0,0.5,1},
                        xticklabels={8,22,34}
                    }
                }
        }
        {
         \ifthenelse{\imodel=1}
        {
         \pgfplotsset{
                    colorbar horizontal,
                    yticklabels={,},
                    point meta min=0,
                    point meta max=1,
                    colorbar style=
                    {
                        at={(0.5,1.15)},
                        anchor=south,
                        width=0.6*\pgfkeysvalueof{/pgfplots/parent axis width},
                        height={5pt},
                        title=$A_\text{EBU}[\text{s}^{-1}]$,
                        xtick={0,1},
                        xticklabels={561,1272}
                    }
                }
        }
        {
         \pgfplotsset{
                    colorbar horizontal,
                    yticklabels={,},
                    point meta min=0,
                    point meta max=1,
                    colorbar style=
                    {
                        at={(0.5,1.15)},
                        anchor=south,
                        width=0.6*\pgfkeysvalueof{/pgfplots/parent axis width},
                        height={5pt},
                        title=$A_\text{TFC}[\text{kg}\text{m}^{-2}\text{s}^{-1}]$,
                        xtick={0,1},
                        xticklabels={0.08,2.7}
                    }
                }
        }
        }
            \begin{axis}
                        [
                            at={($(reference.south west)+(\imodel*55pt,0)$)},
                            anchor=west, 
                            enlargelimits=false,
                            axis equal image,
                            name=last,
                            width=90pt,
                            xtick = {-1,0,1},
                            xmin=-1.5,xmax=1.5,ymin=-0.5,ymax=4,
                        ]
            \end{axis}
    }
\end{tikzpicture}
}
    \caption{Prefactors for the the non-linear and linear reaction rates (Eqs.~(\ref{eq:Arrhenius:nonLinear}), (\ref{eq:EBU:nonlinear}), (\ref{eq:TFC:nonlinear}) and (\ref{eq:LinearizedReactionModels})), determined from the \gls{LES}}
    \label{fig:prefactors}
\end{figure}

The \gls{RANS} combustion models include turbulent statistic quantities, such as the turbulent kinetic energy, $k$, its dissipation rate, $\epsilon$ and the turbulent flame speed $s_\text{t}$. For the above comparison of the different flame models, for the moment they were fixed to constant values along the jet center line. In general, however, these quantities can vary in space as well as in time during the period of a coherent fluctuation and need to be modelled. In this study, the turbulence-flame interactions are assumed not to change in time but spatially. In the \gls{RANS} combustion models, the turbulence-flame interactions are taken into account by lumping the corresponding quantities into the prefactors $A$. Since they are assumed to be constant in time, they can be tuned by fitting the temporal averages of progress variable and reaction rate obtained a priori by the \gls{LES}. Note, however, that although in this way a perfect match with the temporally averaged \gls{LES} reaction rate as shown in Fig.~\ref{fig:reactionRates:nonlinear} can be achieved, this approach cannot improve an inherently weak approach for linearization such as the Arrhenius reaction rate model. Although the temporal mean is corrected by this fitting, the derivation of the combustion model with respect to the state variables appearing in the linear term of the Taylor series in Eqs.~(\ref{eq:Taylor:RANS}) and (\ref{eq:Taylor:DNS}), still describes the slope of the graphs in Fig.~\ref{fig:reactionRates:nonlinear}. In contrast, the prefactor is only a spatially varying amplification. For the sake of consistency and to verify the correctness of reproducing the mean reaction rates, all prefactors, including the Arrhenius prefactor, will be tuned against the \gls{LES} results accordingly. The resulting prefactors will be applied for all remaining results presented in this study. These prefactors of the three flame models are illustrated in Fig.~\ref{fig:prefactors}. 
Note that for illustrative purposes the common logarithm is applied to the prefactor of the Arrhenius type of reaction model, it spans over 24 orders of magnitude. This is due to the exponential term, which is very close to zero for small values of the progress variable. Since the other flame models give already a better estimation of the reaction rate with a constant prefactor, the spatial maximum and minimum of the varying prefactor as determined from the \gls{LES} vary significantly less. The reaction rate based on the \gls{EBU} model already almost coincides with the temporal mean of the \gls{LES} reaction rate (see Fig.~\ref{fig:reactionRates:nonlinear}). The respective prefactor, therefore, is close to constant in almost the entire depicted area, where it takes a value of $A_\text{EBU}\approx 860\, \text{s}^{-1}$. Only at the far upstream region, especially at the flame root, the prefactor of the \gls{EBU} model notably differs from this value.

Unlike the Arrhenius type reaction rate models, \gls{RANS} models provide an approximation of turbulent flame structure, instead of fully resolved flamelets. Nevertheless, extensions of these models are often also applied in \gls{LES}~\cite{Poinsot}. It has been shown, that these derivative LES--EBU/TFC models perform on coarse grids similarly well as Arrhenius type reaction rate models reactions with a thickened flame model~\cite{zhang2019}. As soon as the grid is refined to resolve the actual flame sheet, however, they lead to significantly thicker flame fronts than observed in \gls{DNS}\cite{zhang2019}. As a consequence, the description of flame dynamics at high frequencies based on such models cannot keep up with such based on Arrhenius type reaction rate models. In the approach followed here, however, it is not the goal to describe the dynamics of individual flame sheets. In this study, the focus is on the statistical quantity of a phase average, similar to the Reynolds average addressed in the \gls{RANS} equations. Therefore, the above discussed restriction of the LES--EBU/TFC models does not apply, at least at frequencies for which the hydrodynamic wave length is significantly larger than the laminar flame thickness, which is the case in this study.

The linearized combustion models can be obtained by Eqs.~(\ref{eq:Taylor:DNS}) and (\ref{eq:Taylor:RANS}). Note that in this study, the reaction rates given by Eqs.~(\ref{eq:Arrhenius:nonLinear}), (\ref{eq:EBU:nonlinear}), and (\ref{eq:TFC:nonlinear}) do not depend on the state variables of velocity and pressure, but only on the progress variable. The expressions for the linearized reaction rates are given by
\begin{subequations}
\begin{equation}
    \widehat{\dot{\Omega}}_\text{Ar}  =
    A_\text{Ar} \text{exp} \left(-\frac{T_\text{a}}{\overline{T}}\right) 
    \left( 
        \widehat{\rho} \left( 1- \overline{c}\right)
        - \overline{\rho}  \widehat{c}
        +\overline{\rho} \left( 1- \overline{c} \right) \frac{T_\text{a}}{\overline{T}^2} \widehat{T}
    \right),
\end{equation}
\begin{equation}
    \widehat{\dot{\Omega}}_\text{EBU}  =
    A_\text{EBU} 
    \left(
        \widehat{\rho} \left( \overline{c}- \overline{c}^2\right)
        +\overline{\rho} \left( \widehat{c}- 2\overline{c}\widehat{c}\right)
    \right),\label{eq:linEBU}
\end{equation}
\begin{equation}
    \widehat{\dot{\Omega}}_\text{TFC}  =
    A_\text{TFC} 
 \frac{\nabla \overline{c} \cdot \nabla \widehat{c}}{|\nabla \overline{c}|}.
\end{equation}\label{eq:LinearizedReactionModels}
\end{subequations}

\section{A priori analysis of linearized reaction rates}\label{ch:apriori}
In the previous section, the linearized governing equations for the reacting turbulent flow under investigation were derived and discussed. In this context three different linearized flame models were introduced to offer a description of the turbulent reaction rate $\widehat{\dot{\Omega}}_\text{i}$ (see Eqs.~(\ref{eq:LinearizedReactionModels})).
This section evaluates these linearized flame models. All of the underlying non-linear flame models relate the temporal mean of the reaction rate to the temporal mean of the progress variable, while the linearized flame models correspond to the linear terms of their Taylor expansion around the temporal average of the progress variable. It is, therefore, reasonable to assume that coherent fluctuations in the reaction rate as a result of coherent fluctuations in the progress variable are predicted well by a linearized flame model if the corresponding non-linear model is able to reproduce the shape of the temporally averaged reaction rate obtained from the \gls{LES}. According to this hypothesis, the \gls{EBU} model is expected to provide the best results, followed by the \gls{TFC} model, while the Arrhenius model is expected to give the least accurate results. This hypothesis is addressed in this section.

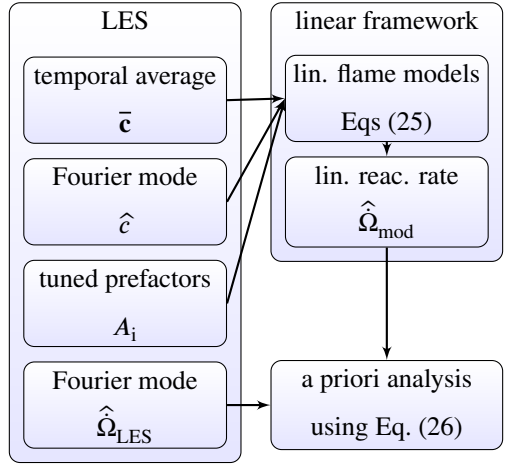
\begin{figure}[t]
    \centering
{\pgfplotsset{
    compat=1.3, 
    every axis/.append style={scale only axis,% axis on top,
    height=2.25cm, width=0.3\textwidth, xmin=0, xmax=2,ymin=0,ymax=1.2
    }
}

\begin{tikzpicture}[node/.style={draw}]
\node (center) {};
%LES
 \node [rounded corners, top color=white, bottom color=blue!10,draw,draw,rectangle, minimum height=0.4\linewidth, minimum width=0.2\linewidth, anchor=north east] (LES) at ($(center)+(-0.09\textwidth,-1.0cm)$) {};
 \node [anchor=north, at=(LES.north)] (LEStext) {LES};
%linearized Framework 
 \node [rounded corners, top color=white, bottom color=blue!10,draw,draw,rectangle, minimum height=0.225\linewidth, minimum width=0.2\linewidth, anchor=north west] (linFram) at ($(LES.north east)+(0.025\textwidth,0.00\textwidth)$) {};
 \node [anchor=north, at=(linFram.north), align = center] (linFramText) {linear framework};

%Fourier modes
 \node [rounded corners, top color=white, bottom color=blue!10,draw,draw,rectangle, minimum height=0.075\linewidth, minimum width=0.175\linewidth, anchor=south] (FourierModes) at ($(LES.south)+(0.0\textwidth,0.0125\textwidth)$) {};
  \node [anchor=north, at=(FourierModes.north), align = center] (FourierModesText) {Fourier mode\\ $\widehat{\dot{\Omega}}_\text{LES}$};
  %tuned prefactors means
 \node [rounded corners, top color=white, bottom color=blue!10,draw,draw,rectangle, minimum height=0.075\linewidth, minimum width=0.175\linewidth, anchor=south] (tunedPrefactor) at ($(FourierModes.north)+(0.0\textwidth,0.0125\textwidth)$) {};
  \node [anchor=north, at=(tunedPrefactor.north), align = center] (tunedPefactorText) {tuned prefactors\\ $A_\text{i}$};
  
   \node [rounded corners, top color=white, bottom color=blue!10,draw,draw,rectangle, minimum height=0.075\linewidth, minimum width=0.175\linewidth, anchor=south] (cfluc) at ($(tunedPrefactor.north)+(0.0\textwidth,0.0125\textwidth)$) {};
  \node [anchor=north, at=(cfluc.north), align = center] (cflucText) {Fourier mode\\ $
  \widehat{c}$};
%temporal means
 \node [rounded corners, top color=white, bottom color=blue!10,draw,draw,rectangle, minimum height=0.075\linewidth, minimum width=0.175\linewidth, anchor=south] (temporalMean) at ($(cfluc.north)+(0.0\textwidth,0.0125\textwidth)$) {};
  \node [anchor=north, at=(temporalMean.north), align = center] (temporalMeanText) {temporal average\\ $\overline{\mathbf{c}}$};

 \node [rounded corners, top color=white, bottom color=blue!10,draw,draw,rectangle, minimum height=0.075\linewidth, minimum width=0.175\linewidth, anchor=south, align = center] (linearResponse) at ($(linFram.south)+(0.0\textwidth,0.0125\textwidth)$) {lin. reac. rate\\ $\widehat{\dot{\Omega}}_\text{mod}$};

  \node [rounded corners, top color=white, bottom color=blue!10,draw,draw,rectangle, minimum height=0.075\linewidth, minimum width=0.175\linewidth, anchor=south] (linEquation) at ($(linearResponse.north)+(0.0\textwidth,0.0125\textwidth)$) {};
  \node [anchor=north, at=(linEquation.north), align = center] (linEquationText) { lin. flame models \\ Eqs~(\ref{eq:LinearizedReactionModels}) };
  
%a priori
 \node [rounded corners, top color=white, bottom color=blue!10,draw,draw,rectangle,
 %minimum height=0.225\linewidth,
 minimum width=0.2\linewidth, anchor= west, align = center] (apriori) at ($(FourierModes.east)+(0.0375\textwidth,0.0\textwidth)$) {a priori analysis \\using Eq.~(\ref{eq:alignment})};
% \node [anchor=north, at=(apriori.north), align = center] (aprioriText) {a priori analysis \\using Eq.~(\ref{eq:alignment})};
  
%temporal means
% \node [rounded corners, top color=white, bottom color=blue!10,draw,draw,rectangle, minimum height=0.075\linewidth, minimum width=0.175\linewidth, anchor=south] (temporalMean) at ($(FourierModes.north)+(0.0\textwidth,0.0125\textwidth)$) {};
 % \node [anchor=north, at=(temporalMean.north)] (temporalMeanText) {temporal average};

    \draw[->,>=latex',thick,draw=black,shorten >=0pt, shorten <=0pt] (temporalMean.east) to (linEquation.west);
    \draw[->,>=latex',thick,draw=black,shorten >=0pt, shorten <=0pt] (tunedPrefactor.east) to (linEquation.west);
    \draw[->,>=latex',thick,draw=black,shorten >=0pt, shorten <=0pt] (cfluc.east) to (linEquation.west);
    \draw[->,>=latex',thick,draw=black,shorten >=0pt, shorten <=0pt] (FourierModes.east) to (apriori.west);
    \draw[->,>=latex',thick,draw=black,shorten >=0pt, shorten <=0pt] (linEquation) to (linearResponse);
    \draw[->,>=latex',thick,draw=black,shorten >=0pt, shorten <=0pt] (linearResponse) to (apriori);
\end{tikzpicture} 

}
    \caption{Schematic illustration of the a priori analysis}
    \label{fig:DiagramAPriori}
\end{figure}

In a \gls{LES} with acoustic forcing, as conducted in this study, the coherent fluctuations of the state variables are known a priori. This allows to perform an a priori analysis, which in a similar way is commonly used in evaluating turbulence models from fully resolved \gls{DNS} data (see e.g. Klein et al.~\cite{Klein2019}). The procedure of this analysis in the framework of this study is sketched in Fig.~\ref{fig:DiagramAPriori}. First, the Fourier modes of the progress variable, $\widehat{c}_\text{LES}$, and the reaction rate, $\widehat{\dot{\Omega}}_\text{LES}$, at the forcing frequency are determined. Subsequently, the Fourier mode of the progress variable, $\widehat{c}_\text{LES}$, is inserted into the linearized flame models of Eqs.~(\ref{eq:LinearizedReactionModels}), to obtain the model flame response $\widehat{\dot{\Omega}}_\text{mod}$. Finally, the Fourier mode of the \gls{LES} reaction rate is compared to the modeled reaction rate using an alignment coefficient defined as 
\begin{equation}
    \mathcal{K} = 
    \frac{\langle \widehat{\dot{\Omega}}_\text{LES} ,\widehat{\dot{\Omega}}_\text{mod}\rangle}
    {
        \norm\big{\widehat{\dot{\Omega}}_\text{LES}}_2 \, 
        \norm\big{\widehat{\dot{\Omega}}_\text{mod}}_2
    },\label{eq:alignment}
\end{equation}
where $\langle\cdot,\cdot\rangle$ describes an Euclidean inner product, while $\norm{\cdot}$ is its associated norm. The alignment factor takes values between $0.0$ and $1.0$, where a value of $1.0$ indicates perfect alignment.

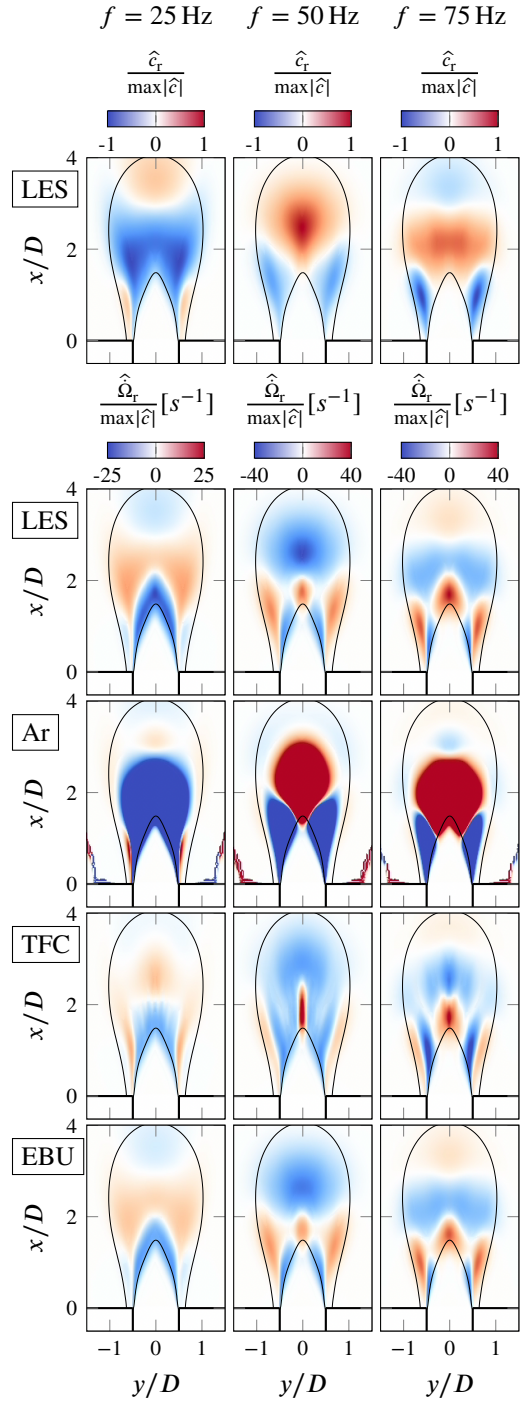
\begin{figure}[hbtp]
    \centering
{\pgfplotsset
{
    compat=1.3, 
    every axis/.append style=
    {
        scale only axis,
        % define the custom colormap
        colormap={my colormap}{
                rgb255=(59, 76, 192),
                rgb255=(255, 255, 255  ),
                rgb255=(180, 4, 38),
        },
        title style={yshift=-4.0pt,},
    }
}
\newlength{\colorbarshift}
\newlength{\colorbarheight}
\setlength{\colorbarheight}{6pt}
\begin{tikzpicture}[scale=1.0]
    \node (reference) {};
    \foreach \ifrequency/\frequency/\limit in {0/25/25,1/50/40,2/75/40}
    {
        \foreach \imodel/\model in {0/cLES,1/LES,2/Ar,3/TFC,4/EBU}
        {
            
            \ifthenelse{\ifrequency = 0}
                    {
                        \ifthenelse{\ifrequency=0}
                        {

                                \pgfplotsset{
                                    ylabel style={align=center},
                                    ylabel={$x/D$},
                                }
                        }
                        {
                        }
                    }
                    {    
                        \pgfplotsset{
                            yticklabels={,},
                        }
                    }
           
            \ifthenelse{\imodel = 4}
                    {
                        \pgfplotsset{
                            xlabel={$y/D$},
                        }
                    }
                    {    
                        \pgfplotsset{
                            xticklabels={,},
                        }
                    }
                    \ifthenelse{\imodel=0}
            {
            }
            {
            \setlength{\colorbarshift}{45.0pt}
            }
            \begin{axis}
                        [
                            at={($(reference.south west)+(\ifrequency*55pt,-\imodel*80pt-\colorbarshift)$)},
                            anchor=west, 
                            enlargelimits=false,
                            axis equal image,
                            name=last,
                            width=90pt,
                            xtick = {-1,0,1},
                            xmin=-1.5,xmax=1.5,ymin=-0.5,ymax=4,
                            xlabel={},
                            ylabel={},
                            xticklabels={},
                            yticklabels={},
                        ]
                        \addplot graphics [xmin=-1.5,xmax=1.5,ymin=-0.5,ymax=4] {pngs/APriori/APriori\frequency Hz\model.png};
                        \addplot [color=black, mark=none,thick] coordinates { (-1.5,0) (-0.5,0)};
                        \addplot [color=black, mark=none,thick] coordinates { (-0.5,0) (-0.5,-0.5)};
                        \addplot [color=black, mark=none,thick] coordinates { (1.5,0) (0.5,0)};
                        \addplot [color=black, mark=none,thick] coordinates { (0.5,0) (0.5,-0.5)};
                         \addplot [color=black, very thin, smooth  ] table [x expr=-\thisrowno{1}/0.02 , y expr = \thisrowno{0}/0.02, col sep=comma]{Data/MeanFlamec002.csv}; 
                        \addplot [color=black, very thin, smooth  ] table [x expr=\thisrowno{1}/0.02 , y expr = \thisrowno{0}/0.02, col sep=comma]{Data/MeanFlamec002.csv}; 
                        \addplot [color=black, very thin, smooth  ] table [x expr=-\thisrowno{1}/0.02 , y expr = \thisrowno{0}/0.02, col sep=comma]{Data/MeanFlamec098.csv}; 
                        \addplot [color=black, very thin, smooth  ] table [x expr=\thisrowno{1}/0.02 , y expr = \thisrowno{0}/0.02, col sep=comma]{Data/MeanFlamec098.csv}; 
            \end{axis}
             \ifthenelse{\imodel=0}
            {
                \pgfplotsset{
                    colorbar horizontal,
                    point meta min=0,
                    point meta max=1,
                    colorbar style=
                    {
                        at={(0.5,1.15)},
                        anchor=south,
                        width=0.7*\pgfkeysvalueof{/pgfplots/parent axis width},
                        height=\colorbarheight,
                        title=$\frac{\widehat{c}_\text{r}}{\text{max}{|\widehat{c}|}}$,
                        xtick={0,0.5,1},
                        xticklabels={-1,0,1}
                    }
                }
            }
            {
            \setlength{\colorbarshift}{45.0pt}
            }
            \ifthenelse{\imodel=1}
            {
                \pgfplotsset{
                    colorbar horizontal,
                    point meta min=0,
                    point meta max=1,
                    colorbar style=
                    {
                        at={(0.5,1.15)},
                        anchor=south,
                        width=0.7*\pgfkeysvalueof{/pgfplots/parent axis width},
                        height=\colorbarheight,
                        title=$\frac{\widehat{\dot{\Omega}}_\text{r}}{\text{max}{|\widehat{c}|}} [s^{-1}]$,
                        xtick={0,0.5,1},
                        xticklabels={-\limit,0,\limit}
                    }
                }
            }
            {
            }
            \begin{axis}
                        [
                            at={($(reference.south west)+(\ifrequency*55pt,-\imodel*80pt-\colorbarshift)$)},
                            anchor=west, 
                            enlargelimits=false,
                            axis equal image,
                            name=last,
                            width=90pt,
                            xtick = {-1,0,1},
                            xmin=-1.5,xmax=1.5,ymin=-0.5,ymax=4,
                        ]
            \end{axis}
            \ifthenelse{\ifrequency = 0}
                    {
                        \ifthenelse{\ifrequency=0}
                        {
                            \ifthenelse{\imodel=0}
                            {
                                \node [anchor=west,draw] (typeText) at ($(last.west)+(-28pt,27pt)$) {LES} ;
                            }
                            {
                                \node [anchor=west,draw] (typeText) at ($(last.west)+(-28pt,27pt)$) {\model} ;
                            }
                        }
                        {
                        }
                    }
                    {
                    }
            \ifthenelse{\imodel=0}
                {
                    \node [anchor=south] (frequencyText) at ($(last.north)+(0pt,45pt)$) {$f=\frequency \, \text{Hz}$} ;
                }
                {
                }
        }
    }
\end{tikzpicture}
}
    \caption{All relevant fields used in the a priori analysis of linearized combustion model. Each column shows the results for one of the forcing frequencies, $f=25\, \text{Hz}$, $f=50\, \text{Hz}$, and $f=75\, \text{Hz}$. The first two rows show the \gls{LES} Fourier modes of progress variable and reaction rate at the respective frequencies. The last three rows show the reaction rates based on the linearized flame models, when the fluctuation in progress variable based on the \gls{LES} (illustrated in the first row) is inserted in the respective equations (Eqs.~\ref{eq:LinearizedReactionModels}).}
    \label{fig:RRApriori}
\end{figure}

Figure~\ref{fig:RRApriori} illustrates the spatial distribution of all fields relevant to the a priori analysis at the three forcing frequencies investigated in the \gls{LES}: $25\,\text{Hz}$ (left column), $50\,\text{Hz}$ (center column), and $75\,\text{Hz}$ (right column). The first row shows the coherent fluctuation in progress variable extracted from the \gls{LES}, which is inserted in Eqs.~(\ref{eq:LinearizedReactionModels}). The second row depicts the coherent fluctuation of the \gls{LES} reaction rate, and is the reference for the linearized combustion models. In the remaining three rows, the modeled reaction rates based on the linearized flame models are depicted. 

The analysis starts with the Arrhenius type reaction rate model, which yields fluctuations far off the charts, with a maximum amplitude of $\text{max}(\widehat{\dot{\Omega}}_\text{Ar})=2.3\cdot 10^{10} \, \text{s}^{-1}$, which occurs mainly in the upstream region of the turbulent flame brush. The flame model appears to react very sensitively to coherent fluctuations of the passive scalar in this region. This can be explained by the high values of the pre-exponential factor there (see Fig.~\ref{fig:prefactors}). It can be concluded, as a consequence, that the error made when ignoring the closure problem in the Arrhenius type reaction rate cannot be compensated by tuning the pre-exponential factor. Note, that the model was also tested for a constant prefactor. In this case, the linearized flame model responded with reaction rates of similar non-realistic nature, however in the downstream region of the turbulent flame brush. Hence, the combustion model based on the Arrhenius type reaction rate and any of its derivatives models, such as the thickened flame model, is not a suitable model for a linear mean field analysis of turbulent flames applying the approach pursuit in this study. 

The second linearized flame model investigated is the \gls{TFC} model (see fourth row in Fig.~\ref{fig:RRApriori}). The magnitude of the reaction rate appears to be in the correct range, much in contrast to the respective results obtained from the Arrhenius type reaction rate model. This can be explained by the fact that the non-linear \gls{TFC} model already gives a significantly better reproduction of the temporal mean reaction rate when a constant prefactor is applied (see Fig.~\ref{fig:reactionRates:nonlinear}). Despite the satisfactory result with respect to the magnitude, the mode shape only remotely resembles the one obtained from the \gls{LES}.

\begin{figure}[t]
    \centering
{\pgfplotsset{
    compat=1.3, 
    every axis/.append style={scale only axis,% axis on top,
    height=2.6cm, width=0.4\linewidth, xmin=20, xmax=80,ymin=0,ymax=1.1,label style={font=\footnotesize},tick label style={font=\footnotesize}
    }
}
    \pgfplotsset{scaled x ticks=false}
\begin{tikzpicture}[scale=1.0000]

\begin{axis}[
scale only axis,
grid=both,
name=real,
ylabel style={align=center},
ylabel={$\text{alignment coefficient}$},
xlabel=$f/\text{Hz}$,
legend pos=south west,
legend cell align={left},
xtick={25,50,75},
    ]

%\addplot [color=black,solid,smooth,  thick] table [x = St , y=TFLES, col sep=comma, solid,color=black]{Data/TFhighNuwJet.csv};% \addlegendentry{$\nu_\text{t,B}$ (Eqn.~\ref{eq:nuSimple})}\label{line:TFLES}
    \addplot coordinates{
        (25,0.991)
        (50,0.995)
        (75,0.9923)
   };   \addlegendentry{EBU};
       \addplot coordinates{
        (25,0.712)
        (50,0.636)
        (75,0.556)
   };   \addlegendentry{TFC};
       \addplot coordinates{
        (25,0)
        (50,0)
        (75,0)
   };
   \addlegendentry{Arrhenius};

\end{axis}
% \begin{axis}[
% scale only axis,
% grid=both,
% anchor=north west,
% at={($(real.north east)+(1.2cm,0)$)},
% name=imag,
% ylabel style={align=center},
% ylabel={$\text{gain}$},
% xlabel=$\text{St}$,
% legend pos=south west,
% legend cell align={left},
% xtick={25,50,75},
%     ]

% %\addplot [color=black,solid,smooth,  thick] table [x = St , y=TFLES, col sep=comma, solid,color=black]{Data/TFhighNuwJet.csv};% \addlegendentry{$\nu_\text{t,B}$ (Eqn.~\ref{eq:nuSimple})}\label{line:TFLES}
%     \addplot coordinates{
%         (25,0.9758)
%   };
%       \addplot coordinates{
%         (25,0.5549)
%   };
%       \addplot coordinates{
%         (25,0)
%   };
%   \addlegendentry{$L/D$=2.5};

% \end{axis}
\end{tikzpicture}
}
    \caption{Alignment coefficient based on Eq.~(\ref{eq:alignment}) as a function of the frequency, illustrating the performance of the linearized combustion models in the a priori analysis}
    \label{fig:alignment}
\end{figure}
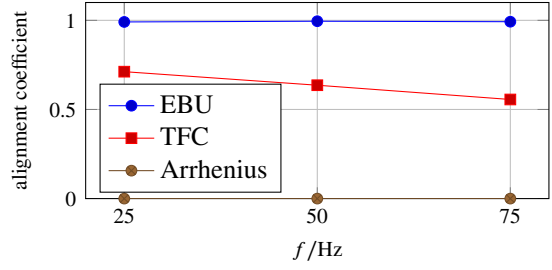

The third combustion model being investigated is the linearized \gls{EBU} model, which performs best with regard to the reproduction of the mean reaction rate, see Section~\ref{ch:linReactionRate}. As expected, Figure~\ref{fig:RRApriori} demonstrates that the reaction rate based on the linearized \gls{EBU} model is in very good agreement with the \gls{LES} results, both in terms of the magnitude and in terms of the mode shape. These findings are underlined by Fig.~\ref{fig:alignment}. It quantifies the results in the a priori study by displaying the alignment coefficient (see Eq.~(\ref{eq:alignment})) of the three linearized combustion models for all three investigated frequencies. The alignment coefficient for the linearized Arrhenius source term is close to zero for all frequencies, while the \gls{TFC} model performs significantly better with values between $0.55$ and $0.7$. Best is as expected the \gls{EBU} model, which yields alignments beyond $0.99$, only. 

The results confirm the above hypothesis, that linear combustion models perform best if their non-linear \gls{RANS} counterpart reproduces the temporal mean reaction rate well. In this study, both the \gls{LES} combustion model, as well as the best suited combustion model, the RANS-\gls{EBU} model, are related in the sense that both are based on a progress variable and include the term $c(1-c)$. If the \gls{EBU}-model is still well suited to investigate flame dynamics, if another combustion model is used in the \gls{LES} is a question of high interest, but reaches beyond the framework of the present study.

\section{A posteriori analysis of the linearized framework}\label{ch:aposteriori}
The a priori analysis identified the proposed \gls{EBU} model of being capable to reproduce the coherent fluctuation in reaction rate when the temporal mean and the coherent fluctuation of the progress variable is provided. In this section, the entire set of linearized equations will be addressed in an a posteriori analysis. Figure~\ref{fig:DiagramAPosteriori} illustrates the underlying approach. The temporal means of velocity and progress variable, the turbulent eddy viscosity and diffusivity determined using Eq.~(\ref{eq:eddyViscosity}), and the tuned prefactor, $A_\text{EBU}$ are determined based on the \gls{LES} results. These fields are used as input to the linearized equations, which then are discretized in the computational domain and solved by the \gls{FELiCS} in the computational domain. The following section describes the numerical strategy and setup of \gls{FELiCS}.
\subsection{Numerical approach in the linearized framework}\label{ch:FELiCS}
The linearized governing Eqs.~(\ref{eq:linear:freqdomain}) and (\ref{eq:linEBU})) constitute a linear set of equations, which can be rewritten as
\begin{equation}
    \mathbf{A} \widehat{\Phi} 
    = \omega \mathbf{B} \widehat{\Phi}
    + f_\text{BC}
    \Rightarrow 
     \widehat{\Phi} = \left(\mathbf{A} - \omega \mathbf{B}\right)^{-1} f_\text{BC}, \label{eq:linSymbol}
\end{equation}
where ${\Phi}=[\mathbf{u},p,c]$, and $f_\text{BC}$ is the load or forcing vector, which arises as a consequence of non-homogeneous Dirichlet boundary conditions. 

The \gls{FELiCS} software takes advantage of the fenics package~\cite{fenics} and is used to discretize this linear set of equations on the computational domain, which due to the rotational symmetry of the \gls{LES} setup can be reduced to two dimensions. Second order, triangular, continuous Galerkin finite elements are used for discretization. The computational mesh is illustrated in Fig.~\ref{fig:FELiCSGrid}. It consists of 20,266 elements. The coarsest elements are of an edge length of $\Delta x = 3\,\text{mm}$. Mesh refinement is performed towards the adiabatic walls in the combustion chamber ($\Delta x = 1.5\,\text{mm}$) in the mixing duct ($\Delta x = 0.8\,\text{mm}$), towards the mixing duct wall ($\Delta x = 0.4\,\text{mm}$), and at the flame root ($\Delta x = 0.25\,\text{mm}$). A mesh convergence study was conducted to demonstrate that results do not change with increasing mesh refinement. The acoustic perturbation is modeled by a Dirichlet boundary condition at the inlet, which fixes the velocity disturbance in streamwise direction to $\widehat{u}_\text{x}=1$. The remaining boundary conditions are listed in Table~\ref{tab:BCs}. After the matrices in Eq.~(\ref{eq:linSymbol}), $A$ and $B$, are discretized in the computational domain using the finite element method, a lower-upper decomposition of the operator $\left(\mathbf{A} - \omega \mathbf{B}\right)$ is performed using SuperLU~\cite{li05AnOverview} in order to solve the linear set of equations given by Eq.~(\ref{eq:linSymbol}).
\begin{figure}[t]
    \centering
{\pgfplotsset{
    compat=1.3, 
    every axis/.append style={scale only axis,% axis on top,
    height=2.25cm, width=0.3\textwidth, xmin=0, xmax=2,ymin=0,ymax=1.2
    }
}

\begin{tikzpicture}[node/.style={draw}]
\node (center) {};
%LES
 \node [rounded corners, top color=white, bottom color=blue!10,draw,draw,rectangle, minimum height=0.4\linewidth, minimum width=0.2\linewidth, anchor=north east] (LES) at ($(center)+(-0.09\textwidth,-1.0cm)$) {};
 \node [anchor=north, at=(LES.north)] (LEStext) {LES};
%linearized Framework 
 \node [rounded corners, top color=white, bottom color=blue!10,draw,draw,rectangle, minimum height=0.3125\linewidth, minimum width=0.2\linewidth, anchor=north west] (linFram) at ($(LES.north east)+(0.025\textwidth,0.00\textwidth)$) {};
 \node [anchor=north, at=(linFram.north), align = center] (linFramText) {linear framework};
 
%Fourier modes
 \node [rounded corners, top color=white, bottom color=blue!10,draw,draw,rectangle, minimum height=0.075\linewidth, minimum width=0.175\linewidth, anchor=south] (FourierModes) at ($(LES.south)+(0.0\textwidth,0.0125\textwidth)$) {};
  \node [anchor=north, at=(FourierModes.north), align = center] (FourierModesText) {Fourier modes\\ $\widehat{\Phi}_\text{LES}$};
  %tuned prefactors means
 \node [rounded corners, top color=white, bottom color=blue!10,draw,draw,rectangle, minimum height=0.075\linewidth, minimum width=0.175\linewidth, anchor=south] (tunedPrefactor) at ($(FourierModes.north)+(0.0\textwidth,0.0125\textwidth)$) {};
  \node [anchor=north, at=(tunedPrefactor.north), align = center] (tunedPefactorText) {tuned prefactor\\ $A_\text{EBU}$};
%temporal means
 \node [rounded corners, top color=white, bottom color=blue!10,draw,draw,rectangle, minimum height=0.075\linewidth, minimum width=0.175\linewidth, anchor=south] (eddyViscosity) at ($(tunedPrefactor.north)+(0.0\textwidth,0.0125\textwidth)$) {};
  \node [anchor=north, at=(eddyViscosity.north), align = center] (eddyViscosityText) {eddy viscosity\\ $\nu_\text{t}\, D_\text{t}$};
 \node [rounded corners, top color=white, bottom color=blue!10,draw,draw,rectangle, minimum height=0.075\linewidth, minimum width=0.175\linewidth, anchor=south] (temporalMean) at ($(eddyViscosity.north)+(0.0\textwidth,0.0125\textwidth)$) {};
  \node [anchor=north, at=(temporalMean.north), align = center] (temporalMeanText) {temporal average\\ $\overline{\mathbf{u}},\,\overline{\mathbf{c}}$};

 \node [rounded corners, top color=white, bottom color=blue!10,draw,draw,rectangle, minimum height=0.075\linewidth, minimum width=0.175\linewidth, anchor=south, align = center] (linearResponse) at ($(linFram.south)+(0.0\textwidth,0.0125\textwidth)$) {linear response\\$\widehat{\Phi}_\text{lin}$};

 \node [rounded corners, top color=white, bottom color=blue!10,draw,draw,rectangle, minimum height=0.075\linewidth, minimum width=0.175\linewidth, anchor=south] (FELiCS) at ($(linearResponse.north)+(0.0\textwidth,0.0125\textwidth)$) {FELiCS};

  \node [rounded corners, top color=white, bottom color=blue!10,draw,draw,rectangle, minimum height=0.075\linewidth, minimum width=0.175\linewidth, anchor=south] (linEquation) at ($(FELiCS.north)+(0.0\textwidth,0.0125\textwidth)$) {};
  \node [anchor=north, at=(linEquation.north), align = center] (linEquationText) { lin. gov. eqs. \\ Eqs~(\ref{eq:linear:freqdomain}) \& (\ref{eq:linEBU}) };
  
%temporal means
% \node [rounded corners, top color=white, bottom color=blue!10,draw,draw,rectangle, minimum height=0.075\linewidth, minimum width=0.175\linewidth, anchor=south] (temporalMean) at ($(FourierModes.north)+(0.0\textwidth,0.0125\textwidth)$) {};
 % \node [anchor=north, at=(temporalMean.north)] (temporalMeanText) {temporal average};

     \draw[->,>=latex',thick,draw=black,shorten >=0pt, shorten <=0pt] (temporalMean) to (FELiCS);
    \draw[->,>=latex',thick,draw=black,shorten >=0pt, shorten <=0pt] (eddyViscosity) to (FELiCS);
     \draw[->,>=latex',thick,draw=black,shorten >=0pt, shorten <=0pt] (tunedPrefactor) to (FELiCS);
     \draw[->,>=latex',thick,draw=black,shorten >=0pt, shorten <=0pt] (linEquation) to (FELiCS);
     \draw[->,>=latex',thick,draw=black,shorten >=0pt, shorten <=0pt] (FELiCS) to (linearResponse);
\end{tikzpicture} 

}
    \caption{Schematic illustration of the a posteriori analysis}
    \label{fig:DiagramAPosteriori}
\end{figure}
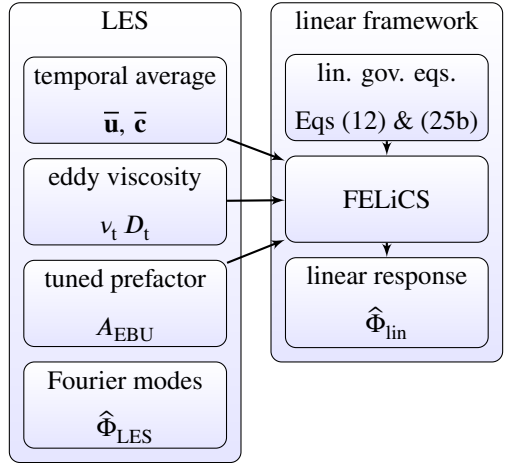
\begin{figure}[t]
    \centering
{\pgfplotsset
{
    compat=1.14, 
    point meta min=0,
    point meta max=1,
    every axis/.append style=
    {
        scale only axis,
        title style={yshift=-4.0pt,},
    },
    height=300pt,
}

\begin{tikzpicture}[scale=1.0]
    \node (reference) {};
                \begin{axis}
                            [
                                anchor=west, 
                                enlargelimits=false,
                                axis equal image,
                                align = center,
                                name=ux,
                                xmin=-1,xmax=20.25,ymin=-0.25,ymax=4.25,
                                xlabel={},
                                ylabel={},
                                xticklabels={,},
                                yticklabels={,},
                            ]
                            \addplot graphics [xmin=-1,xmax=20.25,ymin=-0.25,ymax=4.25] {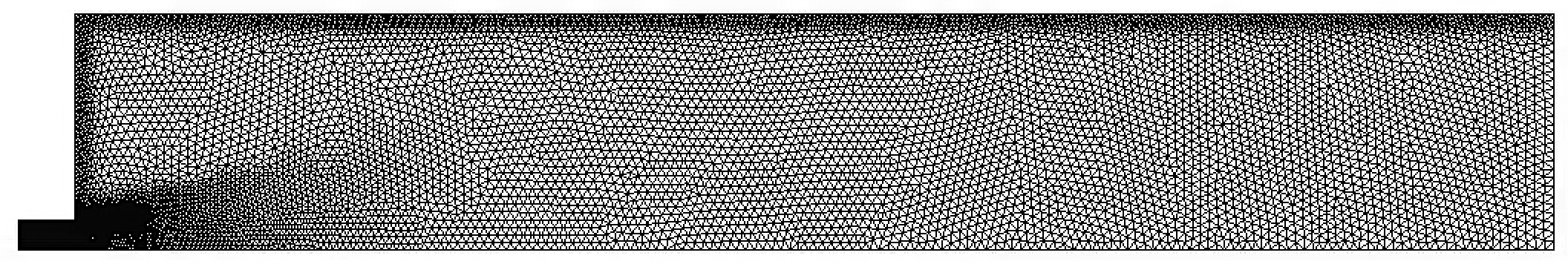};
                            \addplot [color=red, mark=none,line width=2pt] coordinates { (-0.8,0) (-0.8,0.5)};\label{line:inlet}
                            \addplot [color=blue, mark=none,line width=2pt] coordinates { (-0.8,0) (20,0)};\label{line:symmetry}
                            \addplot [color=green, mark=none,line width=2pt] coordinates { (-0.8,0.5) (0,0.5) (0,4) (20,4)};\label{line:wall}
                            \addplot [color=magenta, mark=none,line width=2pt] coordinates { (20,4) (20,0)};\label{line:outlet}
                            %\addplot [color=black, mark=none,thick] coordinates { (-0.5,0) (-0.5,-1.0)};
                            %\addplot [color=black, mark=none,thick] coordinates { (4,0) (0.5,0)};
                            %\addplot [color=black, mark=none,thick] coordinates { (0.5,0) (0.5,-1.0)};
                \end{axis}
                \begin{axis}
                            [
                                anchor=west,
                                at= (ux.west)
                                enlargelimits=false,
                                axis equal image,
                                align = center,
                                name=ux,
                                xmin=-1,xmax=20.25,ymin=-0.25,ymax=4.25,
                                xlabel={$x/D$},
                                ylabel={$r/D$},
                            ]
                \end{axis}
\end{tikzpicture}
}
    \caption{Computational domain of the linearized calculations; computational grid in black, boundaries in color as follows: inlet (\ref{line:inlet}), symmetry (\ref{line:symmetry}), wall (\ref{line:wall}), outlet (\ref{line:outlet})}
    \label{fig:FELiCSGrid}
\end{figure}
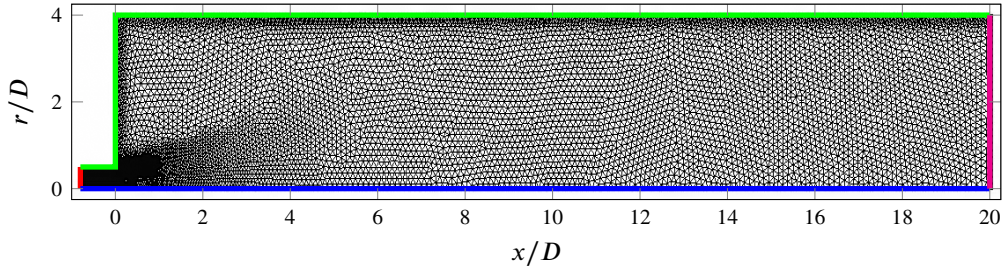

\begin{table}[t]
\centering
\begin{tabular}{|c||c|c|c|c|} 
 \hline
  & Inlet & Symmetry & Wall & Outlet \\  
 \hline\hline
 $\widehat{u}_x$ & $\widehat{u}_x=1$ & $\frac{\partial \widehat{u}_x}{\partial \mathbf{n}}=0$ & $\widehat{u}_x=0$ & $\frac{\partial \widehat{u}_x}{\partial \mathbf{n}}=0$\\ 
 \hline
 $\widehat{u}_r$ & $\widehat{u}_r=0$ & $\widehat{u}_x=0$ & $\widehat{u}_x=0$ & $\frac{\partial \widehat{u}_r}{\partial \mathbf{n}}=0$\\
 \hline
 $\widehat{p}$ & $\frac{\partial \widehat{p}}{\partial \mathbf{n}}=0$ & $\frac{\partial \widehat{p}}{\partial \mathbf{n}}=0$ & $\frac{\partial \widehat{p}}{\partial \mathbf{n}}=0$ & $\widehat{p}=0$\\
 \hline
 $\widehat{c}$ & $\widehat{c}=0$ & $\frac{\partial \widehat{c}}{\partial \mathbf{n}}=0$ & $\frac{\partial \widehat{c}}{\partial \mathbf{n}}=0$ & $\frac{\partial \widehat{c}}{\partial \mathbf{n}}=0$\\
 \hline
\end{tabular}
\caption{Boundary conditions for unknowns in Eq.~\ref{eq:linSymbol} at the boundaries shown in Fig.~\ref{fig:FELiCSGrid}}\label{tab:BCs}
\end{table}
\subsection{Results of the linearized reacting flow equations}\label{ch:linResults}
\begin{figure}[tbhp]
\begin{subfigure}[t]{1.0\textwidth}
{\pgfplotsset
{
    compat=1.3, 
    every axis/.append style=
    {
        scale only axis,
        % define the custom colormap
        colormap={my colormap}{
                rgb255=(59, 76, 192),
                rgb255=(255, 255, 255  ),
                rgb255=(180, 4, 38),
        },
        title style={yshift=-4.0pt,},
    }
}
\newcommand{\limit}{1}
\setlength{\colorbarheight}{6pt}
\begin{tikzpicture}[scale=1.0]
    \node (reference) {};
    \foreach \ifrequency/\frequency/\limit in {0/25/25,1/50/40,2/75/40}
    {
        \foreach \imode/\mode in {0/Imag,1/Magnitude}
        {
            \foreach \isource/\source in {0/LES,1/FELiCS}
            {  
                \ifthenelse{\ifrequency = 0 \AND \isource =0 }
                    {
                        \ifthenelse{\imode=0}
                        {
                            \pgfplotsset{
                                ylabel style={align=center},
                                ylabel={$\widehat{c}_\text{i}$\\$x/D$},
                            }
                        }
                        {
                              \pgfplotsset
                            {
                                ylabel style={align=center},
                                ylabel={$|\widehat{c}|$\\$x/D$},
                            }
                        }
                    }
                    {    
                        \pgfplotsset{
                            yticklabels={,},
                        }
                    }
                \ifthenelse{\imode = 1 }
                    {
                        \pgfplotsset{
                            xlabel={$r/D$},
                            xtick={-1,0,1},
                            xticklabels={1,0,1},
                        }
                    }
                    {    
                        \pgfplotsset{
                            xticklabels={,},
                        }
                    }
                \ifthenelse{\imode=0}
                {
                    \ifthenelse{\isource=0}
                        {
                            \ifthenelse{\ifrequency=0}
                            {
                                \renewcommand{\limit}{0.25}
                            }
                            {
                            }           
                            \ifthenelse{\ifrequency=1}
                            {
                                \renewcommand{\limit}{0.29}
                            }
                            {
                            }
                            \ifthenelse{\ifrequency=2}
                            {
                                \renewcommand{\limit}{0.25}
                            }
                            {
                            }    
                        }
                        {
                        \ifthenelse{\ifrequency=0}
                            {
                                \renewcommand{\limit}{0.25}
                            }
                            {
                            }           
                            \ifthenelse{\ifrequency=1}
                            {
                                \renewcommand{\limit}{0.29}
                            }
                            {
                            }
                            \ifthenelse{\ifrequency=2}
                            {
                                \renewcommand{\limit}{0.25}
                            }
                            {
                            }    
                        }

                }
            {
            }
                \begin{axis}
                            [
                                at={($(reference.south west)+(\ifrequency*120pt+\isource*55pt,-\imode*80pt)$)},
                                anchor=west, 
                                enlargelimits=false,
                                axis equal image,
                                align = center,
                                name=last,
                                width=90pt,
                                xtick = {-1,0,1},
                                xlabel={$y/D$},
                                xmin=-1.5,xmax=1.5,ymin=-0.5,ymax=4,
                            ]
                            \addplot graphics [xmin=-1.5,xmax=1.5,ymin=-0.5,ymax=4] {pngs/APosteriori/\source c\mode \frequency Hz};
                            \addplot [color=black, mark=none,thick] coordinates { (-1.5,0) (-0.5,0)};
                            \addplot [color=black, mark=none,thick] coordinates { (-0.5,0) (-0.5,-0.5)};
                            \addplot [color=black, mark=none,thick] coordinates { (1.5,0) (0.5,0)};
                            \addplot [color=black, mark=none,thick] coordinates { (0.5,0) (0.5,-0.5)};
                             \ifthenelse{\imode=0 \OR \imode =1}
                             {
                                 \addplot [dashed,color=black, very thin, smooth  ] table [x expr=\thisrowno{1}/0.02 , y expr = \thisrowno{0}/0.02, col sep=comma]{Data/neutralline.csv}; 
                                 \addplot [dashed,color=black, very thin, smooth  ] table [x expr=-\thisrowno{1}/0.02 , y expr = \thisrowno{0}/0.02, col sep=comma]{Data/neutralline.csv}; 
                                 \addplot [color=black, very thin, smooth  ] table [x expr=-\thisrowno{1}/0.02 , y expr = \thisrowno{0}/0.02, col sep=comma]{Data/MeanFlamec002.csv}; 
                                 \addplot [color=black, very thin, smooth  ] table [x expr=\thisrowno{1}/0.02 , y expr = \thisrowno{0}/0.02, col sep=comma]{Data/MeanFlamec002.csv}; 
                                 \addplot [color=black, very thin, smooth  ] table [x expr=-\thisrowno{1}/0.02 , y expr = \thisrowno{0}/0.02, col sep=comma]{Data/MeanFlamec098.csv}; 
                                 \addplot [color=black, very thin, smooth  ] table [x expr=\thisrowno{1}/0.02 , y expr = \thisrowno{0}/0.02, col sep=comma]{Data/MeanFlamec098.csv}; 
                             }
                             {
                             }
                \end{axis}
                \ifthenelse{\imode=0}
                {
                    \ifthenelse{\isource =0}
                        {
                            \pgfplotsset
                            {
                                colorbar horizontal,
                                point meta min=0,
                                point meta max=1,
                                colorbar style=
                                {
                                    at={(0.5,1.15)},
                                    anchor=south,
                                    width=0.7*\pgfkeysvalueof{/pgfplots/parent axis width},
                                    height=\colorbarheight,
                                    ylabel style={align=center},
                                    title={$\widehat{c}_\text{LES}$},
                                    xtick={0,0.5,1},
                                    xticklabels={-\limit,0,\limit}
                                }
                            }
                        }
                        {
                            \pgfplotsset
                            {
                                colorbar horizontal,
                                point meta min=0,
                                point meta max=1,
                                colorbar style=
                                {
                                    at={(0.5,1.15)},
                                    anchor=south,
                                    width=0.7*\pgfkeysvalueof{/pgfplots/parent axis width},
                                    height=\colorbarheight,
                                    ylabel style={align=center},
                                    title={$\widehat{c}_\text{lin}$},
                                    xtick={0,0.5,1},
                                    xticklabels={-\limit,0,\limit}
                                }
                            }
                        }
                    }
                    {
                    }
                \begin{axis}
                            [
                                at={($(reference.south west)+(\ifrequency*120pt+\isource*55pt,-\imode*80pt)$)},
                                anchor=west, 
                                enlargelimits=false,
                                axis equal image,
                                align = center,
                                name=lastgrid,
                                width=90pt,
                                xtick = {-1,0,1},
                                xlabel={},
                                xmin=-1.5,xmax=1.5,ymin=-0.5,ymax=4,
                                ylabel={},
                                yticklabels={,},
                                yticklabels={,},
                            ]
                \end{axis}
                \ifthenelse{\imode=0 \AND \isource=0}
                {
                    \node [anchor=south] (frequencyText) at ($(last.north east)+(5pt,35pt)$) {$f=\frequency \, \text{Hz}$} ;
                }
                {
                }
            }
        }
    }
\end{tikzpicture}
}
        \caption{Coherent fluctuation of progress variable; top row: imaginary part, bottom row: magnitude}
    \label{fig:APsosterioric} 
\end{subfigure}
\begin{subfigure}[t]{1.0\textwidth}
\input{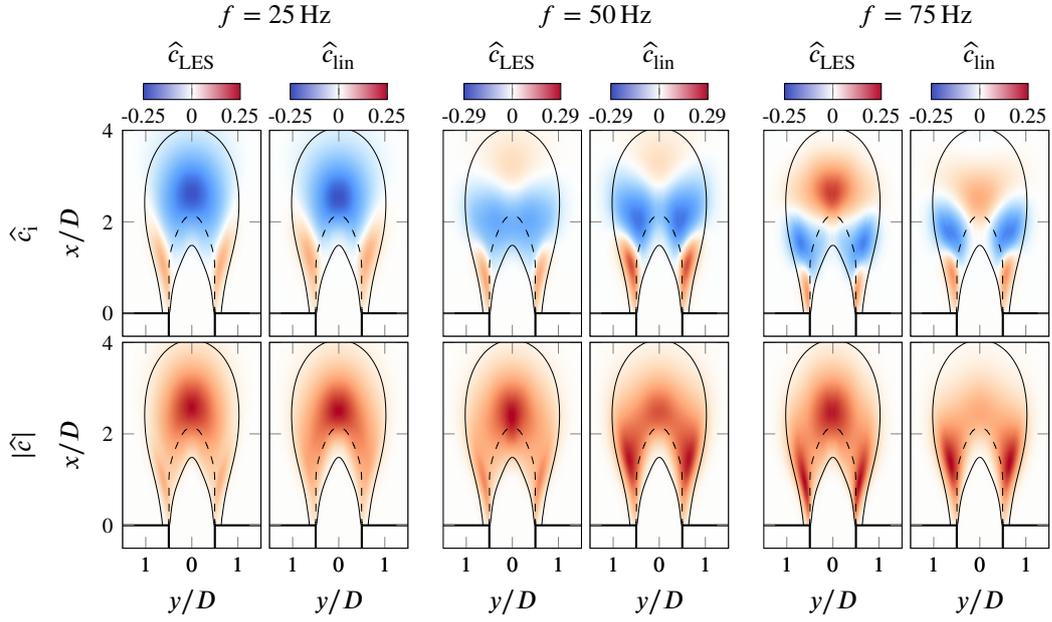}
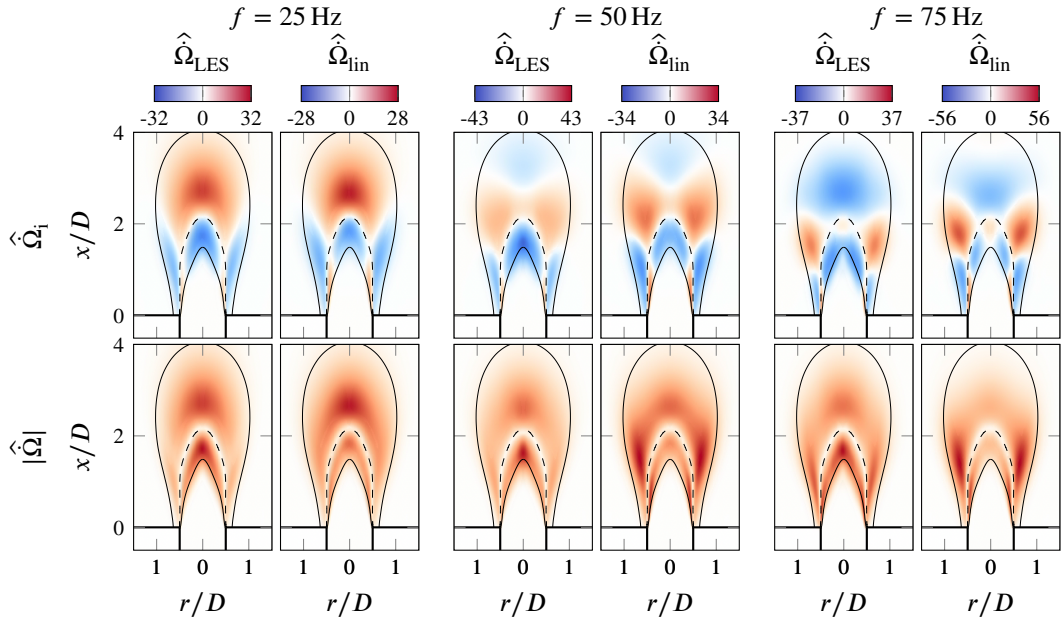
    \caption{Coherent fluctuation of reaction rate; top row: imaginary part, bottom row: magnitude}
    \label{fig:APsosteriorirr} 
\end{subfigure}
\caption{Response of the flame to acoustic perturbations at three frequencies: A posteriori analysis comparing the coherent fluctuations of progress variable and reaction rate based on the linear strategy with the respective fluctuations occurring in the \gls{LES}}\label{fig:APosterioriFlame}
\end{figure}

The response of the flame to acoustic perturbation based on the a posteriori analysis is illustrated in Fig.~\ref{fig:APosterioriFlame}.
Figure~\ref{fig:APsosterioric} illustrates the response of the flame in terms of coherent fluctuations in progress variable for the three frequencies under investigation based on the \gls{LES} and on the linear mean field analysis. The first row displays the imaginary part of the $\widehat{c}$ mode, that is the instantaneous coherent fluctuation at a phase angle of $3/2\,\pi$, while the second row illustrates its magnitude. The columns are divided in three groups of two, each one of the three groups representing one of the frequencies under investigation: $25\, \text{Hz}$, $50\, \text{Hz}$ and $75\, \text{Hz}$. The left plot in each group of columns shows the response in the \gls{LES}, while the right one shows the response based on FELiCS. Overall, the coherent fluctuation in progress variable of the \gls{LES} and the linear mean field analysis are in good agreement. For high frequencies however, slight discrepancies in the mode shape are visible. The phase velocity of the fluctuations seems to be slightly higher in the \gls{LES}, meaning that the linearized approach slightly overestimates the coherent structures wavelengths in streamwise direction. The magnitudes of the fluctuations are shown in the second row of Fig.~\ref{fig:APsosterioric}. While the magnitude of $\widehat{c}$ is predicted with high accuracy by the linear mean field analysis for $f=25\,\text{Hz}$, it appears that this is not the case for the higher frequencies. Here, the magnitude of $\widehat{c}$ at the flame tip is approximately correct, while at the flame root it is overestimated.

The fluctuations in reaction rate are illustrated in Fig.~\ref{fig:APsosteriorirr}. The arrangement of the plots is the same as in Fig.~\ref{fig:APsosterioric}. Again the fields of the instantaneous coherent fluctuation in reaction rate, as can be seen in the first row of the figure, are in very good agreement with the \gls{LES}. Since the very precise prediction of the reaction rate in the a priori analysis and the fact that $\widehat{\dot{\Omega}}_\text{lin}$ is a simple algebraic function of $\widehat{c}_\text{lin}$ (see Eq.~\ref{eq:linEBU}), similar deviations between \gls{LES} and linear results as described above for the coherent fluctuations of the progress variable also occur for the coherent fluctuations in reaction rates. These are best observed in the second row of Fig.~\ref{fig:APsosteriorirr}. Also here, the fluctuations at the flame root are overestimated, while the fluctuations at the flame tip are predicted approximately at the correct magnitude.

Noteworthy about both the \gls{LES} as well as the linear responses in reaction rate is a neutral line, where $\widehat{\dot{\Omega}}$ appears to be close to zero. This neutral line reaches from one burner edge through the flame tip back to the other burner edge. The occurrence of this region of low coherent fluctuations in reaction rate can be explained by an analysis of the linearized combustion model (Eq.~\ref{eq:linEBU}). Using Eqs.~(\ref{eq:rhoOfCLin}) and assuming that the prefactor, $A_\text{EBU}$, is approximately constant, it can be expressed as a function of the temporal mean fields and the coherent fluctuation in progress variable, reading
\begin{equation}
    \widehat{\dot{\Omega}}_\text{EBU}  =
    \underbrace{
    A_\text{EBU} 
    \left(
        - \frac{\overline{\rho}}{\overline{T}}\left(T_\text{b} - T_\text{b} \right)\left( \overline{c}- \overline{c}^2\right)
        +\overline{\rho} \left( 1 - 2\overline{c}\right)
    \right)
    }_\text{$S(\overline{c})$}
    \widehat{c}.  \label{eq:linEBU:sensitivity}
\end{equation}
The expression before $\widehat{c}$ on the right hand side of Eq.~(\ref{eq:linEBU:sensitivity}) is constant in time. Therefore, this term can be seen as the spatially varying sensitivity, $S$, of the the coherent fluctuations in reaction rate with respect to $\widehat{c}$.  
\begin{figure}[t]
    \centering
\input{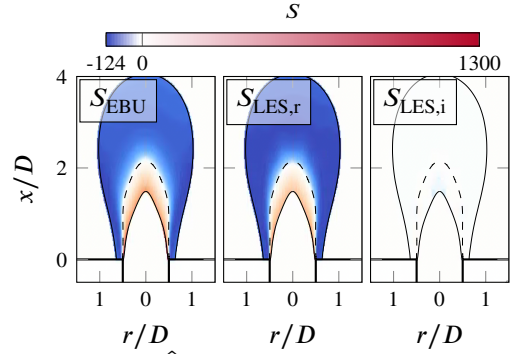}
    \caption{Sensitivity of the fluctuation in reaction rate, $\widehat{\dot{\Omega}}$, with respect to the fluctuation in progress variable, $\widehat{c}$; left: based on the EBU model with fitted prefactor based on Eq.~\ref{eq:linEBU:sensitivity}; center/right: real and imaginary part of the sensitivity based on the \gls{LES} (Eq.~\ref{eq:sensitivityLES})}
    \label{fig:sensitivities}
\end{figure}
This quantity is illustrated in Fig.~\ref{fig:sensitivities}. The left plot shows the sensitivity as obtained from Eq.~(\ref{eq:linEBU:sensitivity}). Based on the linearized \gls{EBU} model, the reaction rate fluctuation as a consequence of a perturbation in the progress variable depends on the location in the flame. While coherent fluctuation in reaction rate and coherent fluctuation in progress variable are directly proportional in the upstream region ($\overline{c}\lesssim 0.28$) of the flame the opposite is the case in the downstream region ($\overline{c}\gtrsim 0.28$). The model predicts a neutral line, where $S=0$, located approximately at $\overline{c}\approx 0.28$, which indicates that a fluctuation in progress variable has no impact on the reaction rate, and therefore the fluctuation in the reaction rate is zero. In the plots of the coherent fluctuations of the reaction rate in Fig.~\ref{fig:APsosteriorirr}, the iso-contour of $\overline{c}_0$ is given by the dashed line. It appears that it is in good agreement with the location of zero reaction rate, giving an additional validation of the strategy of using the linearized \gls{EBU} model.

The sensitivity can be furthermore obtained empirically using the Fourier modes at the forcing frequencies of the \gls{LES} with acoustic forcing by
\begin{equation}
    S_\text{LES} = \frac{\widehat{\dot{\Omega}}_\text{LES}}{\widehat{c}_\text{LES}}. \label{eq:sensitivityLES}
\end{equation}
Since Fourier modes are used in Eq.~(\ref{eq:sensitivityLES}), the resulting sensitivity is complex. Its real part is illustrated exemplary for the forcing frequency of $25\,\text{Hz}$ in the center of Fig.~\ref{fig:sensitivities}, while its imaginary part is shown on the right hand side. The real part of the sensitivity based on the \gls{LES} is in very good agreement with the modeled sensitivity. Furthermore, the imaginary part is very close to zero indicating that there is no significant time delay between the fluctuations in progress variable and in reaction rate. Overall the agreement between the modelled sensitivity and the sensitivity obtained from the \gls{LES} with acoustic forcing demonstrates that the flame dynamics are reproduced reasonably well by the linearized \gls{EBU} model, which is in line with the alignment coefficient.

\begin{figure}[tbhp]
    \centering
\input{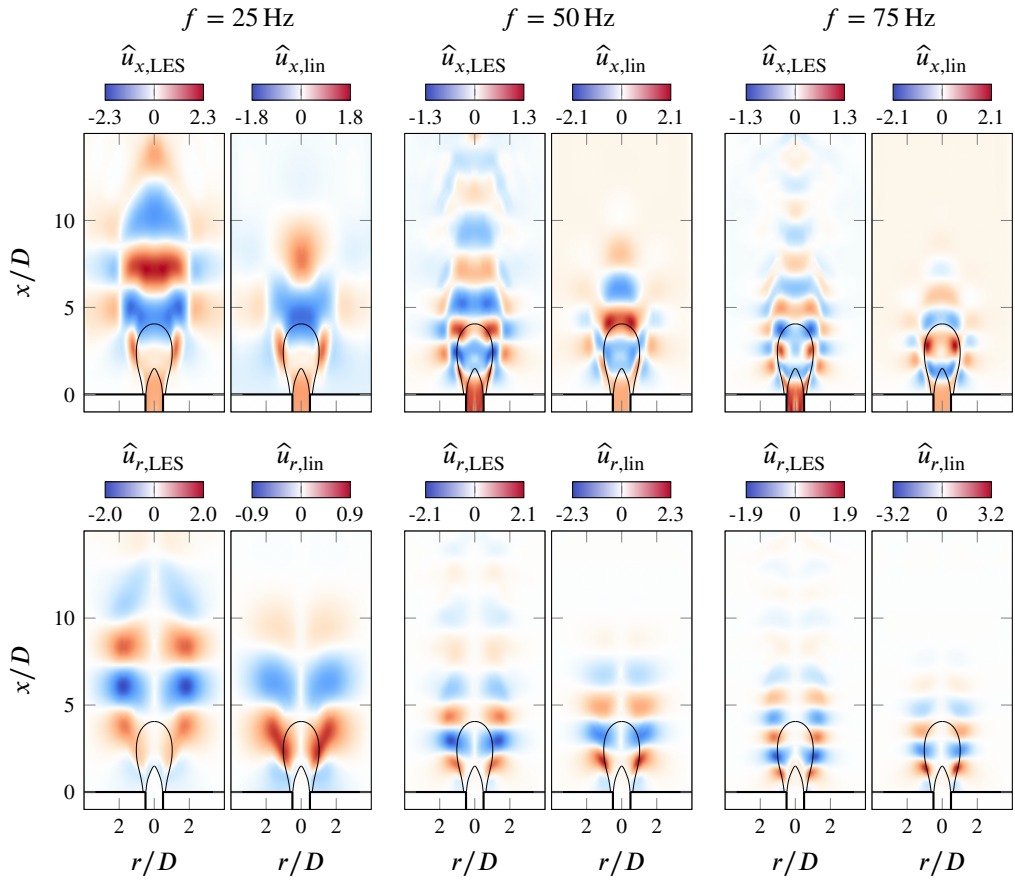}
    \caption{Velocity fluctuations in streamwise (top row) and radial direction (bottom row) as a result to acoustic forcing at $f=25\, \text{Hz}$ (left column pair), $f=25\, \text{Hz}$ (center column pair); In each column pair, \gls{LES} Fourier modes are shown on the left and linear results are shown on the right.}
    \label{fig:FigureAPosterioriUx}
\end{figure}

Finally, the instantaneous responses in velocities to upstream acoustic forcing are illustrated in Fig~\ref{fig:FigureAPosterioriUx}. Here, the real part of the velocity fluctuations in streamwise (upper row) and radial (lower row) direction are shown, indicating the response of the flow at a phase angle of $0$ of the acoustic perturbation. The \gls{LES} Fourier modes show the characteristic shape of Kelvin--Helmholtz vortex rings, as investigated recently by Casel et al.~\cite{Casel2022}. It appears that for low frequencies, the velocity fluctuations are the strongest downstream of the turbulent flame brush, while for intermediate and high frequencies the respective maximum occurs within the flame. Furthermore, the decay rate of the modes in streamwise direction increases with the frequency, leading to more compact mode shapes at the higher frequencies. Both of these behaviours are well captured by the linear approach. It appears however that the coherent velocity fluctuations are decaying at a lower rate in the \gls{LES} in comparison with the linear results. Besides this, the linear analysis appears to overestimate the magnitude of the velocity perturbations for the higher frequencies, which explains the deviations in the coherent fluctuations of progress variable and reaction rate in this region and at these frequencies. Considering the significant simplifications made when deriving the linearized governing equations, the overall agreement in the a posteriori analysis is very satisfactory.

\section{Conclusion}\label{ch:conclusion}
The goal of the present study is to develop a strategy for linear mean field analysis of a turbulent reacting flow with an active flame approach taking into account coherent fluctuations in heat release. It is worked out that non-linear combustion models which are capable of reproducing the temporal mean reaction rates when fed with the temporal mean state variables are suitable for the linear mean field analysis. Generally \gls{RANS} combustion models fulfill this condition. We substantiate this claim by performing an a priori analysis of three different combustion models, with the anticipated result: the \gls{RANS}-\gls{EBU} model, which gives the best description of the temporal mean reaction rate, is best suited for the linearized approach. The temporal mean values of the state variables, which are needed as input to the linear analysis, were provided by \gls{LES} to avoid problems connected with the computation of the mean fields of reacting flows with the help of \gls{RANS}. This approach suggests the general procedure when performing mean field analysis of a turbulent reacting flow is applying a linearized \gls{RANS} combustion model to the temporal means of a high fidelity simulations such as \gls{LES}. The application of different combustion models in the \gls{LES} and testing the results against the above condition, remains for future investigation. One alternative to this could be a future development of a combustion model, which is designed specifically for the linearized framework. This approach could take advantage of the fact, that the statistics of the turbulent flame dynamics are a priori known from the \gls{LES} or \gls{DNS}.

Additionally to the a priori analysis, the response of the turbulent flame to acoustic perturbations is investigated in the linear mean filed analysis, which is based on the full set of linearized reacting flow equations including the linearized \gls{EBU} model. A comparison of the resulting response modes with the respective Fourier modes obtained from the \gls{LES} (labelled a posteriori analysis) yields favourable results. The flow response to acoustic actuation is governed by a dominant Kelvin--Helmholtz vortex roll-up. The linear results predict this behavior. With respect to the mode shapes, there is good agreement between the linear results and the \gls{LES}. One main difference, however, is an overestimation of the magnitude and phase velocity of the Kelvin--Helmholtz vortex rings, which develop downstream of the burner edge, especially for the higher frequencies. Since the interaction of these velocity fluctuations with the turbulent flame can be seen as the main driver of reaction rate fluctuations, also the magnitude and phase velocity of the reaction rate is overestimated in the a posteriori analysis. Since the a priori analysis demonstrated the high accuracy of the linearized flame model, these discrepancies seem not to be caused by a model error in the linearized reaction rate, but by different model assumptions. One reason for these differences could be the assumption of frozen turbulence, meaning that eddy dissipation is derived from the unforced \gls{LES} and it is assumed that it does not vary during the period of an oscillation. In this context, an a priori study, focusing on the correct modelling of the turbulence closure in the linearized framework could remedy the situation.

The principles for linearization of a flame model which were laid out in this study constitute a well founded strategy on how to proceed when choosing a linear flame model. They can be applied in the future to further develop linearized turbulent flame models for the purpose of studying various kinds of dynamics of turbulent reacting flows, such as thermoacoustic instabilities, intrinsic thermoacoustic instabilities as well as direct and indirect combustion noise. In this context, possible points of connection in future studies are a generalization to non-adiabatic conditions or an investigation of turbulence-flame interaction and flame noise using a resolvent analysis. Furthermore, once compressibility is taken into account, a thermoacoustic feedback cycle could be investigated using the full set of linearized equations.  

\section*{Acknowledgements}
Funded by the Deutsche Forschungsgemeinschaft (DFG – German Research Foundation) under the project number 441269395.

\newpage
\bibliography{mybibfile}
\newpage
\end{document}